\documentclass[acmlarge]{acmart}

\usepackage[utf8]{inputenc}
\usepackage[T1]{fontenc}
\usepackage{graphicx}
\usepackage{adjustbox}
\usepackage{tabularx}
\usepackage{subcaption}
\usepackage[english]{babel}
\usepackage[figuresright]{rotating}
\usepackage{makecell}
\usepackage{array}
\usepackage{caption}
\usepackage{multirow}
\usepackage{url}
\usepackage{float}
\usepackage{hyperref}
\usepackage{adjustbox}
\usepackage{rotating} 
\usepackage{array}
\usepackage{makecell}
\usepackage{tikz}
\usepackage{afterpage}
\usepackage{longtable,lscape}
\usepackage{enumitem}
\usepackage{color, colortbl}
\usepackage{wrapfig}
\definecolor{Gray}{gray}{0.9}
\definecolor{LightGray}{gray}{0.6}
\usepackage[most]{tcolorbox}
\hypersetup{colorlinks, citecolor={black}, filecolor=black, linkcolor=black, urlcolor=black}

\definecolor{red}{rgb}{0, 0, 0}
\newcommand{\olsi}[1]{\,\overline{\!{#1}}} 
\AtBeginDocument{%
  \providecommand\BibTeX{{%
    \normalfont B\kern-0.5em{\scshape i\kern-0.25em b}\kern-0.8em\TeX}}}

\setcopyright{acmcopyright}
\copyrightyear{2020}
\acmYear{2020}
\acmDOI{00.0000/0000000.0000000}
\acmJournal{IMWUT}
\acmVolume{00}
\acmNumber{0}
\acmArticle{000}
\acmMonth{0}

\setcopyright{none}

\begin{document}

\title{Examining the Social Context of Alcohol Drinking in Young Adults with Smartphone Sensing}


\author{Lakmal Meegahapola}
\email{lmeegahapola@idiap.ch}
\affiliation{
  \institution{Idiap Research Institute \& EPFL, Switzerland}
}

\author{Florian Labhart}
\email{florian@idiap.ch}
\affiliation{\institution{Idiap Research Institute, Switzerland}}

\author{Thanh-Trung Phan}
\email{trungphan@idiap.ch}
\affiliation{\institution{Idiap Research Institute, Switzerland}}

\author{Daniel Gatica-Perez}
\email{gatica@idiap.ch}
\affiliation{\institution{Idiap Research Institute \& EPFL, Switzerland}}


\renewcommand{\shortauthors}{Meegahapola et al.}

\begin{abstract}

According to prior work, the type of relationship between a person consuming alcohol and others in the surrounding (friends, family, spouse, etc.), and the number of those people (alone, with one person, with a group) are related to many aspects of alcohol consumption, such as the drinking amount, location, motives, and mood. Even though the social context is recognized as an important aspect that influences the drinking behavior of young adults in alcohol research, relatively little work has been conducted in smartphone sensing research on this topic. In this study, we analyze the weekend nightlife drinking behavior of 241 young adults in a European country, using a dataset consisting of self-reports and passive smartphone sensing data over a period of three months. Using multiple statistical analyses, we show that features from modalities such as accelerometer, location, application usage, bluetooth, and proximity could be informative about different social contexts of drinking. We define and evaluate seven social context inference tasks using smartphone sensing data, obtaining accuracies of the range 75\%-86\% in four two-class and three three-class inferences. Further, we discuss the possibility of identifying the sex composition of a group of friends using smartphone sensor data with accuracies over 70\%. The results are encouraging towards supporting future interventions on alcohol consumption that incorporate users’ social context more meaningfully and reducing the need for user self-reports when creating drink logs for self-tracking tools and public health studies.

\end{abstract}



\maketitle

\section{Introduction}\label{sec:introduction}

In western countries, alcohol consumption is a leading risk factor for mortality and morbidity \cite{Kuntsche2013}. The consumption of several drinks in a row, commonly referred as binge drinking or heavy drinking, can lead to many short-term adverse consequences not only for the person drinking (e.g., unprotected sex, injury, accidents or blackouts \cite{Labhart2018}) but also at the family and community levels (e.g. violence, drunk driving \cite{Laslett2011, Kuntsche2017}). On a larger time frame, heavy alcohol consumption can also lead to long-term consequences, such as poor academic achievement, diminished work capacity, alcohol dependence and premature death \cite{Marshall2014}. Adolescence and early adulthood appear as a particularly critical period of life for the development of risky alcohol-related behaviors since heavy alcohol consumption in late adolescence appears to persist into adulthood \cite{White2013}. In order to limit excessive drinking among adolescents and young adults, it is essential to understand the etiology and antecedents of drinking occasions \cite{Kuntsche2005}. Prior work in social and epidemiological research on alcohol has emphasized the importance of the social context in shaping people’s alcohol use and motives \cite{McCarty1985, Freisthler2014, Kuntsche2005} in the sense that the consumption of alcohol or not, and the amounts consumed, vary depending on the presence or absence of family members \cite{Poelen2007, VanDerVorst2006, VanDerVorst2005, Peterson1994}, of friends or colleagues \cite{Mohr2001,Osgood2013,Poelen2007,Dick2007}, and of the spouse or partner \cite{Leonard2003,Leadley2000,Lisco2012}. Additionally, a recent literature review showed that although the type of company is generally not a significant direct predictor of alcohol-related harm, young adults tend to experience more harm, independent of increased consumption, when they drink in larger groups \cite{stevely_drinking_2020}. 

Recent developments in ambulatory assessment methods (i.e., the collection of data in almost real time, e.g. every hour, and in the participant’s natural environment \cite{Shiffman2008,Smyth2003}) using smartphones made it possible to assess the type and the number of people present over the course of real-life drinking occasions \cite{kuntsche_future_2014, Labhart2021}. Compared to cross-sectional retrospective surveys traditionally used in alcohol epidemiological and psychological research, this type of approach allows to capture the interplay between drinking behaviors and contextual characteristics at the drinking event level in more detail \cite{Kuntsche2014}. For instance, evidence shows that larger numbers of drinking companions are associated with increased drinking amounts over the course of an evening or night \cite{thrul_impact_2015,smit_drinking_2015}, and that this relationship is mediated by the companions’ gender \cite{Thrul2017}. By repetitively collecting information from the same individuals over multiple occasions, ambulatory assessment methods are able to capture a large diversity of social contexts of real-life drinking occasions (e.g. romantic date with a partner, large party with many friends, family dinner) with the advantages of being free of recall bias and of participants serving as their own controls.

In addition to the possibility of capturing in-situ self-reports, smartphone-based apps have the potential to provide just-in-time adaptive interventions (JITAI) and feedback \cite{NahumShani2016, Meegahapola2021}. Feedback systems primarily rely on identifying users’ internal state or the context that they are in, to offer interventions or support (feedback) \cite{Kumar2013, SpruijtMetz2014}. Leveraging these ideas, recent studies in alcohol research have used mobile apps to provide interventions to reduce alcohol consumption using questionnaires, self-monitoring, and location-based interventions \cite{Gustafson2014, Attwood2017, You2015}. Furthermore, mobile sensing research has used passive sensing data from wearables and smartphones to infer aspects that could be useful in feedback systems, such as inferring drinking nights \cite{Santani2018}, inferring non-drinking, drinking, and heavy drinking episodes \cite{Bae2017}, identifying walking under alcohol influence \cite{Kao2012}, and detecting drunk driving \cite{Dai2010}. Hence, given that the characteristics of the social context have been identified as central elements of any drinking event, it appears as a central target for inferring drinking occasions. However, to the best of our knowledge, mobile sensing has not been widely used to automatically infer the social context of alcohol drinking events. To further understand the importance of identifying social context using mobile sensing, consider the following example. If an app could infer a heavy-drinking episode (as shown by \cite{Bae2017}), it could provide an intervention. However, there is a significant difference between drinking heavily alone or with a group of friends \cite{skrzynski_systematic_2020, gmel_who_2008}. Drinking several drinks in a row alone might indicate that the person is in emotional pain or stressed (also known as "coping" drinking motive) \cite{Kuntsche2005, skrzynski_systematic_2020}. However, drinking several drinks is common when young adults go for a night-out with friends \cite{gmel_who_2008}. In a realistic setting, for a mobile health app to provide useful interventions or feedback, the knowledge of the social context, in addition to knowing that the user is in a heavy-drinking episode, could be vital. Hence, understanding fine-grained contextual aspects related to alcohol consumption using passive sensing is important, and could also open new doors in mobile interventions and feedback systems for alcohol research.

Further, there are a plethora of alcohol tracking, food tracking, and self-tracking applications in app stores, that primarily rely on user self-reports \cite{Meyer2020, Meegahapola2020, Santani2018}. Even though gaining a holistic understanding regarding eating or drinking behavior is impossible without capturing contextual aspects regarding such behaviors, prior work has shown that people tend to reduce the usage of apps that require a large number of self-reports, and tend to use health and well-being applications that function passively \cite{Meegahapola2021}. Mobile sensing offers the opportunity to infer attributes that otherwise require user self-reports, hence reducing user burden \cite{Meegahapola2021, Meegahapola2021v2, Meegahapola2020}. In addition, mobile sensing could infer attributes to facilitate search acceleration in food/drink logging apps \cite{Jung2020}. The social context of drinking alcohol is a variable that could benefit from smartphone sensing in an alcohol tracking application. As a whole, the idea of using smartphone sensing, in addition to capturing self-reports, is to gain a holistic understanding regarding the user context passively, that could otherwise take a long time-span if collected using self-reports. Considering all these aspects, We address the following research questions:

\begin{itemize}[wide, labelwidth=!, labelindent=0pt]
    \item[\textbf{RQ1:}] What social contexts around drinking events can be observed by analyzing self-reports and smartphone sensing data corresponding to weekend drinking episodes  of a group of young adults?
    \item[\textbf{RQ2:}] Can young adults' social context of drinking be inferred using sensing data? What are the features that are useful in making such inferences?
    \item[\textbf{RQ3:}] Are social context inference models robust to different group sizes? Can mobile sensing features infer the sex composition (same-sex, mixed-sex, opposite-sex), when drinking is done in a group of friends or colleagues? 
\end{itemize}{}

By addressing the above research questions, our work makes the following contributions: 
\begin{itemize}[wide, labelwidth=!, labelindent=0pt]
    \item[\textbf{Contribution 1:}] Using a fine-grained mobile sensing dataset that captures drinking event level data from 241 young adults in a European country, we first show that there are differences in self-reporting behavior among men and women, regarding drinking events done with family members and with groups of friend/colleagues. Next, using various statistical techniques, we show that features coming from modalities such as accelerometer, location, bluetooth, proximity, and application usage are informative regarding different social contexts around which alcohol is consumed. 
    \item[\textbf{Contribution 2:}] We first define seven social context types, based on the number of people in groups (e.g., alone, with another person, with one or more people, with two or more people) and the relationship between the participant and others in the group (e.g., family or relatives, friends or colleagues, spouse or partner). Then, based on the above context types, we evaluate four two-class and three three-class inference tasks regarding the social context of drinking, using different machine learning models, obtaining accuracies between 75\% and 86\%, with all passive smartphone sensing data. In addition, we show that models that only take inputs from single sensor modalities such as accelerometer and application usage, could still perform reasonably well across all seven social context inferences, providing accuracies over 70\%.
    \item[\textbf{Contribution 3:}] For the specific case of drinking with friends or colleagues, we show that mobile sensor data could infer the sex composition of groups (i.e. same-sex, mixed-sex, or opposite-sex) in a three-class inference task, obtaining an accuracy of 75.8\%. 
\end{itemize}{}

The paper is organized as follows. In Section~\ref{sub:related_work}, we describe the background and related work. In Section~\ref{sec:mobile_app}, we describe the study design, data collection procedure, and feature extraction techniques. In Section~\ref{sec:data_analysis} and Section~\ref{sec:statistical_analysis}, we present a descriptive analysis and a statistical analysis of dataset features. We define and evaluate inference tasks in Section~\ref{sec:inference}. Finally, we discuss the main findings in Section~\ref{sec:inference}, and  conclude the paper in Section~\ref{sec:discussion}.

\section{Background and Related Work}\label{sub:related_work}

\subsection{The Social Context of Drinking Alcohol}\label{subsec:alcohol_perspective}

While there are numerous definitions for the term social context in different disciplines, in this paper, we borrow the concept commonly used in alcohol research \cite{labhart_individual_2014, cox_systematic_2019, Kuntsche2005, McCarty1985}, which refers to either one or both of the following aspects: (1) \textit{type of relationship}: the relationship between an individual and the people in the individual’s environment with whom she or he is engaging, and (2) \textit{number of people}: the number of people belonging to each type of relationship, with whom the individual is engaging. By combining the two aspects, a holistic understanding of the social context of drinking of an individual can be attained.

The consumption of alcohol is associated with different contextual characteristics. These characteristics include the type of setting (e.g., drinking location), its physical attributes (e.g., light, temperature, furniture), its social attributes (e.g., type, size, and sex-composition of the drinking group, on-going activities), and the user’s attitudes and cognition \cite{McCarty1985}. Applied to real-life situations, this conception underlines the changing nature of the drinking context, in the sense that the variety of situations during which alcohol might be consumed is rather large. For instance, across three consecutive days, the same person might drink in a restaurant during a date with a romantic partner, join a large party at a night club with many attendees, and finally, join a quiet family dinner at home. 

Among all contextual characteristics, the composition of the social context is a central element of any drinking occasion, since the consumption of alcohol is predominantly a social activity for non-problematic drinkers \cite{skrzynski_systematic_2020}. Among adolescents and young adults, previous literature has shown that amounts of alcohol consumed on any specific drinking occasion vary depending on the type and number of people present \cite{cox_systematic_2019}. The type of relationship that received the most attention so far is the presence of friends, in terms of number and of sex composition. Converging evidence shows that the likelihood of drinking \cite{bersamin_identifying_2016} and drinking amounts are positively associated with the size of the drinking group \cite{smit_drinking_2015, thrul_impact_2015, gallupe_adolescent_2013}. Unfortunately, the group size is generally used as a continuous variable, preventing the identification of a threshold at which the odds of drinking in general or drinking heavily increase. Evidence regarding the sex composition of the group, however, provided mixed results, with some studies indicating that more alcohol is consumed in mixed-sex groups \cite{Thrul2017, lipperman-kreda_social_2018} while others suggesting that this might rather be the case in same-sex groups \cite{ander_2017}. The influence of the presence of the partner (e.g. boyfriend or girlfriend) within a larger drinking group has not been investigated, but evidence suggests that alcohol is less likely to be consumed and in lower amounts in a couple situation (i.e., the presence of the partner only) \cite{Thrul2017, jackson_contextual_2016}. It should be noted that these studies only suggest correlational links between the contextual characteristics and drinking behaviors and should not be interpreted as causal relationships.

The presence or absence of members of the family also play an important role in shaping adolescents and  young adults' drinking behaviors. In particular, the presence of parents and their attitude towards drinking are often described as being either limiting or facilitating factors, but evidence in this respect is inconclusive. For instance, the absence of parental supervision was found to be associated to an increased risk for drinking at outdoor locations and young adults' home \cite{thrul_associations_2018} suggesting that their presence might decrease this risk. However, another study shows that parents’ knowledge about the happening of a party is negatively associated with the presence of alcohol, but there was no relationship between whether a parent was present at the time of the party and the presence of alcohol \cite{friese_teen_2014}. Lastly, parents might also facilitate the use of alcohol by supplying it, especially to underage drinkers \cite{jackson_contextual_2016, gilligan_parental_2012}. To sum up, evidence on the impact of the presence or absence of parents on young people’s drinking appears mixed as this might be related to their attitude towards drinking, with some parents being more tolerant or strict than others \cite{pape_drinking_2017}. Lastly, it should be noted that the presence of siblings has rarely been investigated, but unless they have a supervision role in the absence of parents, their role within the drinking group might be similar to one of friends.

\subsection{Alcohol Consumption and Mobile Phones}\label{subsec:alcohol_mobile}

\subsubsection{Mobile Apps for Interventions in Alcohol Research}

Many mobile apps in alcohol research focus on providing interventions or feedback to users to reduce alcohol consumption \cite{Crane2018, Davies2017, Gustafson2014}. Crane et al. \cite{Crane2018} conducted a randomized controlled trial using the app called "Drink Less", to provide interventions. This app relied on user self-reports, and they concluded that the app helped reduce alcohol consumption. Moreover, Davies et al. \cite{Davies2017} conducted a randomized controlled trial with an app called "Drinks Meter", that provided personalized feedback regarding drinking. This app also used self-reports to provide feedback. Similarly, many mobile health applications in alcohol research that provide users with interventions or feedback, primarily used self-reports \cite{Hides2018, Donnell2019,Carra2016}. Regarding sensing, Gustafson et al. \cite{Gustafson2014} deployed an intervention app called ACHESS, that provided computer-based cognitive behavioral therapy and additional links to useful websites, and this app provided interventions to users when they entered pre-defined high-risk zones, primarily relying on location sensing capabilities of the smartphone. LBMI-A \cite{Dulin2013} by Dulin et al. is another study that is similar to ACHESS. As a summary, alcohol epidemiology research that used mobile apps primarily targeted interventions based on self-reports or simple sensing mechanisms. Even though many studies have identified that self-reports are reasonably accurate to capture alcohol consumption amounts \cite{Lintonen2004}, studies have also stated that heavy-drinking episodes are often under-reported when self-reporting \cite{Northcote2011}. In addition, unless there is a strong reason for users to self-report, there is always the risk of users losing motivation to use the app over time.

\subsubsection{Smartphone Sensing for Health and Well-Being}

Smartphones allow sensing health and well-being aspects via continuous and interaction sensing techniques, both of which are generally called passive sensing \cite{Meegahapola2021}. This capability has been used in areas such as stress \cite{Can2019,Lu2012}, mood \cite{LiKamWa2013, Spathis2019}, depression \cite{Wang2018, Canzian2015}, well-being \cite{Lin2012, Lane2014}, and eating behavior \cite{Biel2018, Meegahapola2020, Meegahapola2021v2}. If we consider drinking related research in mobile sensing, Bae et al. \cite{Bae2017} conducted an experiment with 30 young adults for 28 days, and used smartphone sensor data to infer non-drinking, drinking, and heavy-drinking episodes with an accuracy of 96.6\%. They highlighted the possibility of using such inferences to provide timely interventions. Santani et al. \cite{Santani2018} deployed a mobile sensing application among 241 young adults for a period of 3 months, to collect sensor data around weekend nightlife events. They showed that sensor features could infer drinking and non-drinking nights with an accuracy of 76.6\%. Kao et al. \cite{Kao2012} proposed a phone-based system to detect feature anomalies of walking under the influence of alcohol. Further, Arnold et al. \cite{Arnold2015} deployed a mobile application called Alco Gait, to classify the number of drinks consumed by a user into sober (0-2 drinks), tipsy (3-6 drinks) or drunk (greater than 6 drinks) using gait data, obtaining reasonable accuracies. While most of these studies focused on detecting drinking events/episodes/nights, we focus on inferring the social contexts of drinking events. 

\subsubsection{Event Detection and Event Characterization in Mobile Sensing}

Smartphone sensing deployments can be classified into two based on the study goal \cite{Meegahapola2021}: (a) Event Detection (e.g. drinking alcohol, eating food, smoking, etc.) and (b) Event Characterization (characteristics of the context that helps understand the event better -- e.g. social context, concurrent activities, ambiance, location, etc.). For domains such as eating behavior, there are studies regarding both event detection (identifying eating events \cite{Bedri2017, Morshed2020}, inferring meal or snack episodes \cite{Biel2018}, inferring food categories \cite{Meegahapola2020v2}) and event characterization (inferring the social context around eating events \cite{Meegahapola2020}). Inferring mood \cite{Rodriguez2017, LiKamWa2013} as well as identifying contexts around specific moods \cite{Darvariu2020} has been attempted in ubicomp. However, even though alcohol epidemiology researchers have attempted to characterize alcohol consumption to gain a more fine-grained understanding about drinking, mobile sensing research has not been focused on the social context aspect thus far, even though some studies have looked into event detection \cite{Bae2017, Santani2018, Kao2012, Arnold2015}. Hence, we aim to address this research gap by focusing on the social context of drinking alcohol using smartphone sensing.

\section{Data, Features, and Tasks}\label{sec:mobile_app}

\subsection{Mobile Application, Self-Reports, and Passive Sensing}\label{subsec:mobile_app_sensing_reports}

We obtained a dataset regarding young adults' nightlife drinking behavior, from our previous work \cite{Santani2018}. This dataset contains smartphone sensor data and self-reports regarding the drinking behavior of a 241 young adults (53\% men) in Switzerland, during weekend nights, throughout a period of three months, and was collected as a collaboration between alcohol researchers, behavioral scientists, and computer scientists. In this section, we briefly describe the study design, the data collection procedure, and feature extraction technique. A full description regarding the ethical approval, deployment, and the data collection procedure can be found in \cite{YouthAtNight2021, Santani_ACMMM_2015, Santani2018}.

\textbf{Mobile App Deployment.}\label{subsec:mobile_app} To collect data from study participants, an android mobile application was deployed, and this app had two main components: \textbf{(a) Drink Logger:} used to collect in-situ self-reports during weekend nights (Friday and Saturday nights, from 8.00pm to 4.00am next day). The app sent notifications hourly, asking whether users wanted to report a new drink; and \textbf{(b) Sensor Logger:} used many passive sensing modalities to collect data, including both continuous (accelerometer, battery, bluetooth, location, wifi) and interaction (applications, screen usage) sensing. The application was deployed from September to December 2014. The study participants were young adults with ages ranging from 16 to 25 years old (mean=19.4 years old, SD=2.5). More details regarding the deployment can be found in \cite{Santani2018}.

\textbf{Self-Reports.}\label{subsec:self_reports} Whenever they were about to drink an alcoholic or non-alcoholic drink, participants were requested to take a picture of it and to describe its characteristics and the drinking context using a series of self-reported questionnaires \cite{labhart_capturing_2020}. Participants labeled the drink type using a list of 6 alcoholic drinks (e.g. beer, wine, spirits, etc.) and 6 non-alcoholic drinks (e.g. water, soda, coffee, etc.). Then, in accordance with the definition of social context we adopted in Section~\ref{subsec:alcohol_perspective}, participants reported the type and number of people present for each of the following categories: (a) partner or spouse; (b) family or relatives; (c) male friends or colleagues; (d) female friends or colleagues; and (e) other people (called \textit{type of relationship} in the remainder of the paper). These five categories were adopted from prior work in alcohol research \cite{labhart_individual_2014}. Next, for each type of relationship, participants reported the \textit{number of people} using a 12-point scale with 1-point increments from 0 to 10, plus ‘more than 10’; with the exception of partner or spouse which could either be absent (coded as 0) or present (1). This scale was designed to measure variations in the social context, following the assumption that the presence of each person counts within small groups, but that the additional value of each extra person is less important within larger groups (e.g. 10 or more people). Further, information about participants including age, sex, occupation, education level, and accommodation were collected in a baseline questionnaire. Overall, by selecting self-reports of situations when participants reported the consumption of an alcoholic drink, we were left with 1254 self-reports for the analysis.

\textbf{Passive Smartphone Sensing.}\label{subsec:passive_sensing} To gain a fine-grained understanding about users' drinking behavior, passive sensing data were collected during the same time period when participants self-reported alcohol consumption events. The chosen sensing modalities were Accelerometer (ACC), Applications (APP), Location (LOC), Screen (SCR), Battery (BAT), Bluetooth (BLU), Wifi (WIF), and Proximity (PRO). A dataset summary is given in Table~\ref{table:allfeatures} and an extensive description is given in \cite{Santani2018, Phan2020}.

\begin{table*}
\centering
\caption{Summary of features extracted from mobile sensors (134). Sensor data are aggregated for every 10-minute time slot from 8pm to 4am. For all the given features, average, minimum, and maximum were calculated during the matching phase, hence resulting in a total of 402 sensing features for each alcohol consumption event.}
\label{table:allfeatures}
\resizebox{\textwidth}{!}{%
\begin{tabular}{>{\arraybackslash}m{4.2cm} >{\arraybackslash}m{16cm}}

\rowcolor{red!5}
\textbf{Sensor} & 
{Sensor Description} 
\\

\rowcolor{red!5}
{-- Feature Type (\# of features)} & 
{Feature Description} 
\\ 

\hline

\rowcolor{gray!15}
\textbf{Location} &
Location data were continuously collected for a time period of 1-minute during each 2-minute time slot. Collected data included data source, longitude, latitude, signal strength, and accuracy. 
\\

\rowcolor{gray!5}
-- Attributes (10) & 
\{min., max., med., avg., std.\} of avg. of speed and sensor accuracy 
\\ 

\rowcolor{gray!5}
-- Signal (3) & 
3 signal strengths (GPS, network, unknown)
\\

\rowcolor{gray!15}
\textbf{Accelerometer} & 
Values from all three axes of the sensor were collected, 10 seconds continuously, at a frequency of 50Hz, during every minute. we calculated (a) basic statistics from raw sensor data from the X, Y, and Z-axes~\cite{Santani2018}; (b) aggregated statistics related to acceleration (m, mNew, dm) and signal magnitude area (mSMA) by combining data from three axes \cite{Karantonis2006, Mathie2004, Phan2020}; and (c) angle between acceleration and the gravity vector \cite{Phan2020, Santani2018}.
\\

-- Raw (15) & 
\{min., max., med., avg., std.\} of avg. of xAxis, yAxis, zAxis of accelerometer
\\ 

-- Angle (15) &
\{min., max., med., avg., std.\} of angle of xAxis, yAxis, zAxis with g vector
\\ 

-- Dynamic (20) &
\{min., max., med., avg., std.\} of mSMA, dm, m, mNew values
\\ 


\rowcolor{gray!15}
\textbf{Bluetooth} &
The list of available devices was captured as Bluetooth logs, once every 5 minutes. Features such as the number of devices around, signal strengths, and empty scan counts were captured.
\\ 
 
\rowcolor{gray!5}
-- Count (4) & number of bluetooth IDs surrounding devices, records, bluetooth scan count, empty scan count
\\ 

\rowcolor{gray!5}
-- Strength (5) &
\{min., max., med., avg., std.\} of bluetooth strength signal of surrounding devices
\\

\rowcolor{gray!15}
\textbf{Wifi} &
The list of available devices was captured as WiFi logs, once every 5 minutes. Features such as the number of hotspots around, signal strengths, and the empty scan counts were captured.
\\ 

-- Count (2) &
wifi record , wifi id set
\\ 

-- Attributes (10) &
\{min., max., med., avg., std.\} of  level, frequency of wifi hotspot
\\

\rowcolor{gray!15}
\textbf{Application} &
Applications were categorized into 33 groups (e.g., art \& design, food \& drink, social, games, etc.) based on the categorization provided in google play store \cite{Google2021}. Using the categories and running apps, statistics such as the total number of running apps and the number of running apps based on categories were calculated \cite{LiKamWa2013, Santani2018, Phan2020}.
\\ 

\rowcolor{gray!5} 
-- Count (2) &
app count, app record
\\ 

\rowcolor{gray!5}
-- Category (33) &
normalized 33-bin histogram of 33 application categories 
\\

\rowcolor{gray!15}
\textbf{Proximity} &
Several statistical featured were derived using raw values of the proximity sensor. 
\\ 

-- Count (1)  &
proximity records
\\ 

-- Distance (5) &
\{min., max., med., avg., std.\} of distance from phone to objects
\\

\rowcolor{gray!15}
\textbf{Battery} &
Battery levels were captured once every five minutes, and status changes were captured whenever a change occured. Hence, several features including battery full, discharging, charging, battery level, and whether the phone is plugged-in or not were derived. 
\\ 

\rowcolor{gray!5}
-- Status (5) &
5 battery statuses
\\ 

\rowcolor{gray!5}
-- Level (5) &
\{min., max., med., avg., std.\} of battery level
\\ 

\rowcolor{gray!5} 
-- Count (2) &
count of battery records and plugged times 
\\

\rowcolor{gray!15}
\textbf{Screen} &
Screen data were recorded whenever the screen status changed. Using the captured data, we derived the percentage of screen-on time. 
\\ 

-- Usage (1)  &
percentage of screen on time
\\
\hline

    \end{tabular}
    }
\end{table*}

\subsection{Aggregation and Matching of Self-Reports and Passive Sensing Data}\label{sec:aggregation}

Prior studies that used this dataset primarily considered \textit{user-night} as the point of analysis (e.g., inferring nights of alcohol consumption vs. no alcohol consumption \cite{Santani2018}, inferring heavy-drinking nights \cite{Phan2020}, etc.). However, in this study, we consider drink-level data, that is more fine-grained. We prepared the self-report dataset such that each entry corresponds to a drinking event. Then, to combine sensor data and self-reports in a meaningful manner, we used the following two-phase technique, that was adopted from prior ubicomp research \cite{Bae2018, Rodriguez2017, Meegahapola2020, Biel2018}: 

\textbf{Phase 1 (Aggregation)}: We aggregate raw sensor data for every ten-minute window throughout the night. Different techniques were used for the aggregation for sensors. Hence, for a user-night, we have 48 ten-minute windows, from 8.00pm to 4.00am next day. For each feature derived from each sensor, we have 48 values (6 ten-minute windows per hour X eight hours per night) for a user-night. For instance, if there is a feature F$_{1}$ derived from sensor S$_{1}$, for each for each user and for each night, F$_{1}$ would have 48 values, that represent time windows from 8.00pm-8.09pm, 8.10pm-8.19pm, 8.20pm-8.29pm, until 03.50am-03.59am of next day.

\begin{figure*}[t]
\begin{center}
    \begin{minipage}[t]{0.55\textwidth}
        \includegraphics[width=\textwidth]{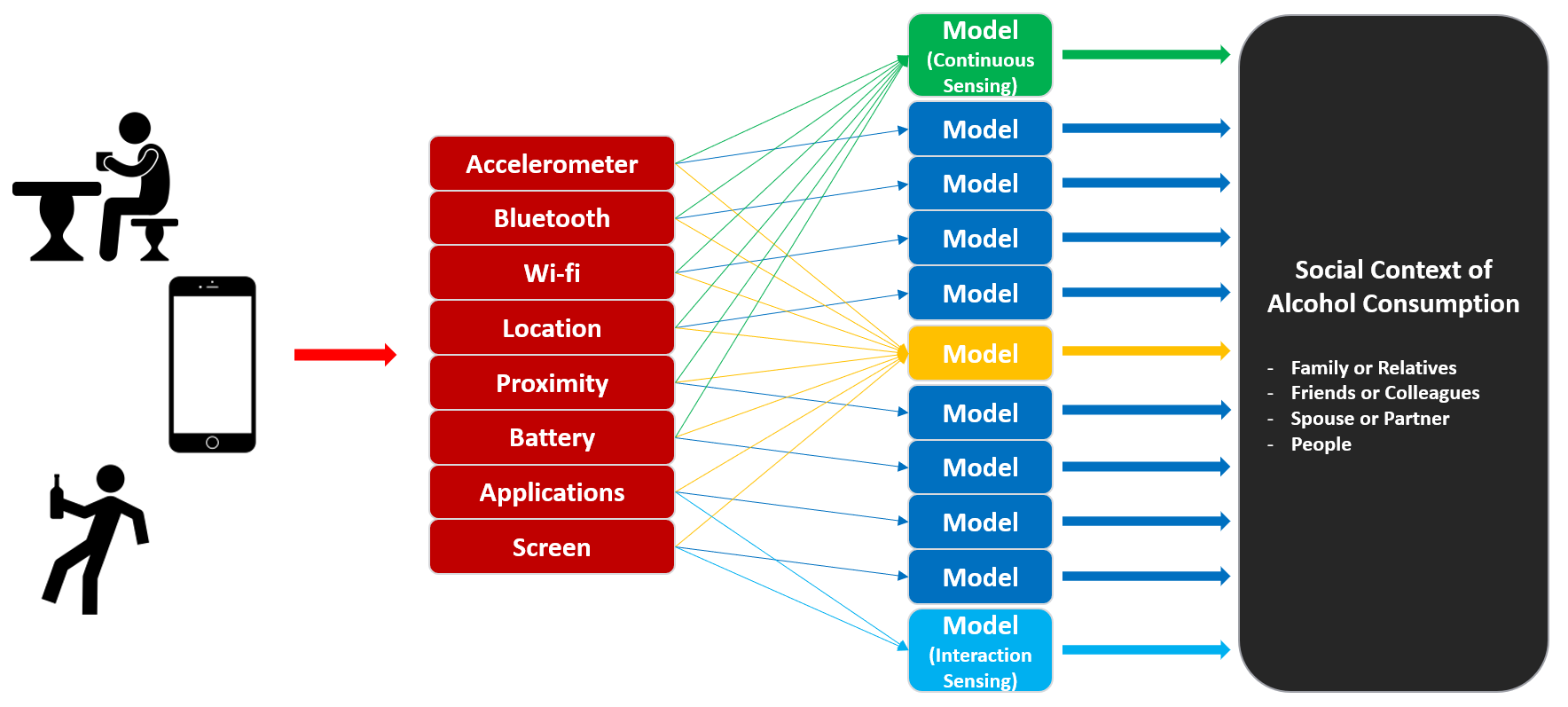}
        \caption{A schematic diagram representing the summary of the study}
        \label{fig:block_diagram}
    \end{minipage}
     \hfill
     \begin{minipage}[t]{0.44\textwidth}
        \includegraphics[width=\textwidth]{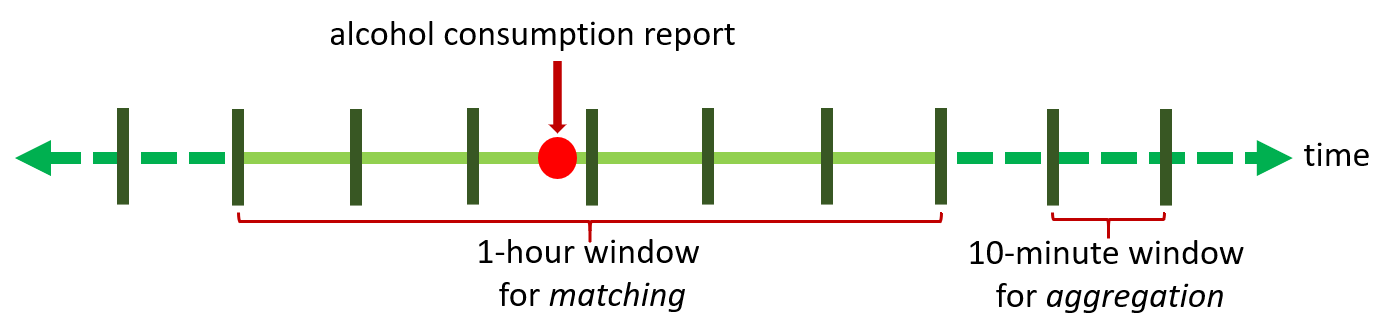}
            \caption{Diagram summarizing the two-phase technique for combining self-reports and sensing data}
        \label{fig:aggregation_matching}
    \end{minipage}
\end{center}
\vspace{-0.2 in}
\end{figure*}

\textbf{Phase 2 (Matching)}: During this phase, features are matched to alcohol consumption self-reports using a one-hour window (approximately, from 30 minutes before the alcohol consumption self-report, to 30 minutes after the drinking self-report). For instance, if the drinking was reported at 10.08pm, we calculate the average (\_avg), maximum (\_max), and minimum (\_min) values for each feature using values corresponding to six ten-minute windows (obtained in Phase 1) from 9.40pm to 10.39pm, and match those value to the self-report. The idea behind this aggregation is to capture sensor data around drinking events, expecting that contextual cues could be informative of different social contexts. 
The summary of passive sensing features is provided in Table~\ref{table:allfeatures}. These features were derived for every ten-minute time slot throughout the night, as discussed in Phase 1 (Aggregation), resulting in a total of 134 features. Then, in accordance with the procedure in Phase 2 (Matching), a one-hour time window around each drinking self-report was considered. By considering the average, minimum, and maximum of the six values corresponding to the one-hour time window, 402 passive sensing features were included for each alcohol drinking self-report in the final dataset. This two-phase technique is summarized in Figure~\ref{fig:aggregation_matching}. After following this technique for all self-reports and sensor features, we obtained a dataset. We removed data points with incomplete sensor data (152 records with unavailable sensor data, mainly location, wifi, or bluetooth data), not enough data for the matching phase (102 records, for drinking events that were done between 8.00pm-8.30pm and 3.30am-4.00am), and self-reports that were produced while visiting other countries (59 records, when the participant traveled while being in the study). The final dataset contained 941 complete drinking reports with sensor features.

\subsection{Deriving Two-Class and Three-Class Social Context Features}\label{sub:social_context} 

In Section~\ref{sub:related_work}, we described how social contexts such as being with/without family members, friends/colleagues, and spouse/partner could be associated with drinking behavior. In addition, under Section~\ref{subsec:self_reports}, we described the type of social contexts reported by participants. Among them, features such as \textit{with male friends/colleagues}, \textit{with female friends/colleagues}, and \textit{with family members} had twelve-point scales, and \textit{with partner/spouse} had a two-point scale. However, for the purpose of this analysis, we reduced the twelve-point scale to low-dimensional scales (two-point and three-point), with the objective of capturing social context group dynamics, that are meaningful in terms of drinking events such as: being alone, with another person, or in a group of two or more. We followed the following steps. 

First, except for the feature \textit{with partner or spouse} which is already two-class, we minimized the scale of other features to two-classes and three-classes. For two-class features, the values could be either zero or one, whereas -- \textbf{zero:} the participant is not with anyone belonging to the specific social context; and \textbf{one:} the participant is with one or more others belonging to the specific social context (hence, in a group). For three-class features, the values could be either zero, one, or two as follows -- \textbf{zero:} the participant is not with anyone belonging to the specific social context; \textbf{one:} the participant is with one other person belonging to the specific social context (hence, in a group of two people); \textbf{two:} the participant is with two or more people belonging to the specific social context (hence, in a larger group).

\noindent Then, we derived several new features using the existing features: \begin{itemize}
    \item \textit{without friends/colleagues vs. with friends/colleagues} (two-class): this aggregated features about men and women friends/colleagues into a single two-class variable by discarding the sex demographic attribute of friends/colleagues.
    \item \textit{without friends/colleagues vs. with another friend/colleague vs. with two/more friends/colleagues} (three-class): this aggregated features about the men and women friends/colleagues into a single three-class feature.
    \item \textit{without people vs. with people} (two-class): this feature combines all the two-class social contexts to estimate the overall two-class social context of the user.
    \item \textit{without people vs. with another person vs. with two/more people} (three-class): this feature combines all the other three-class social contexts and the two-class feature \textit{with partner/spouse}, to estimate the overall three-class social context of the user.
\end{itemize}

The final set of social context features used for this study are summarized in Table~\ref{tab:social_contexts}. In accordance with the definition of social context proposed in Section~\ref{sub:social_context}, these features capture two aspects. First, they capture the relationships between the study participant and people engaging with the participant during alcohol consumption. Second, they capture group dynamics for each relationship (e.g., alone, with another person -- small group of two people, with two/more people -- comparatively large group, etc.). According to prior work in alcohol research, both perspectives are important to obtain a fine-grained understanding about drinking behavior \cite{McCarty1985,Beck2008}. The summary of our analytical setting is presented in Figure~\ref{fig:block_diagram}. 

\begin{table}[tb]
    \caption{Summary of social contexts in the final dataset.}
    \label{tab:social_contexts}
    \centering
    \resizebox{0.7\textwidth}{!}{%
    \begin{tabular}{l l l}

    \rowcolor{red!5}
    \textbf{Social Context} & 
    \textbf{Classes} & 
    \textbf{Interpretation} 
    \\ [0.5ex] 
    \hline
     
    family$_{two}$ &
    2 &
    without vs. with one/more family members/relatives \\ 
    
    \rowcolor{gray!5}
    partner$_{two}$ &
    2 & 
    without vs. with the partner/spouse \\

    friends$_{two}$ &
    2 & 
    without vs. with one/more friends/colleagues \\
    
    \rowcolor{gray!5}
    people$_{two}$ &
    2 & 
    without vs. with one/more people \\
    
    \cmidrule{1-3}
    
    family$_{three}$ &
    3 &
    without vs. with one vs. with two/more family members/relatives \\ 

    \rowcolor{gray!5}
    friends$_{three}$ &
    3 & 
    without vs. with one vs. with two/more friends/colleagues \\
    
    people$_{three}$ &
    3 & 
    without vs. with one vs. with two/more people \\
    
    \hline 
    \end{tabular}
    }
\end{table}

\section{Descriptive Analysis (RQ1)}\label{sec:data_analysis}

\begin{figure*}[t]
\begin{center}
    \begin{minipage}[t]{0.49\textwidth}
        \centering
        \includegraphics[width=\textwidth]{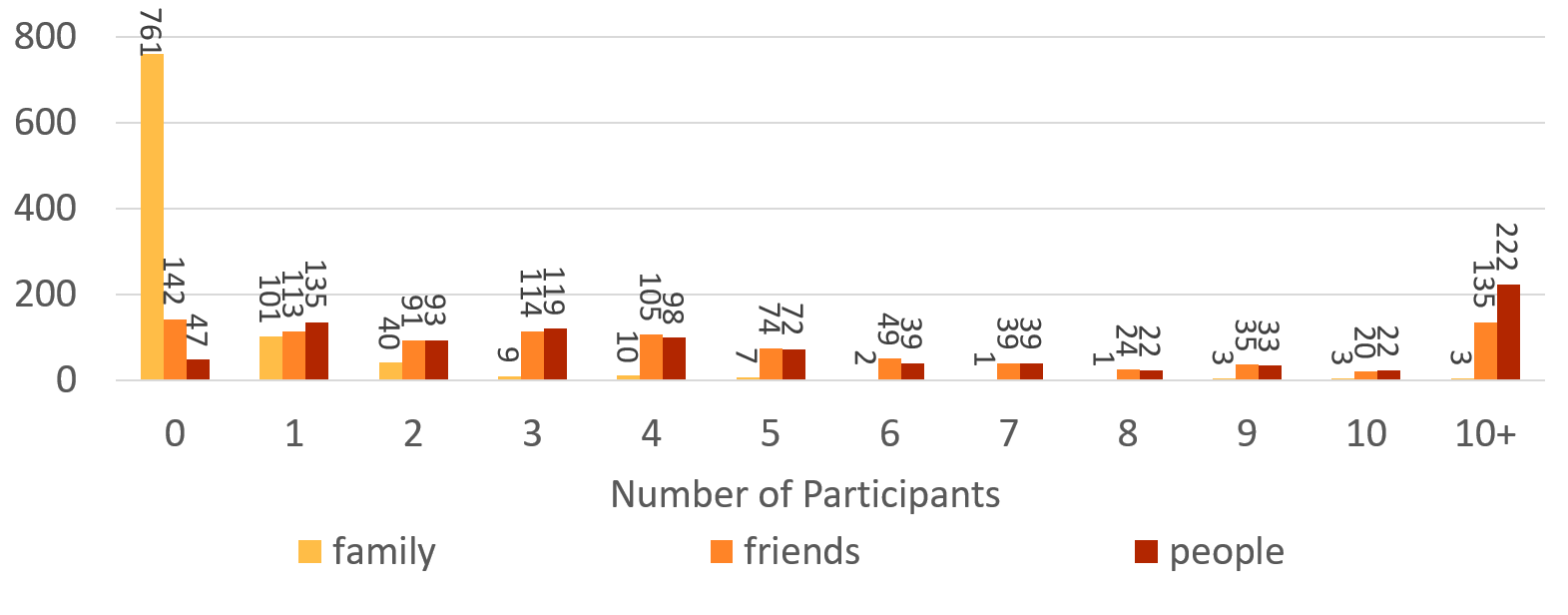}
        \caption{Distribution of original self-report features (family, friends) and a derived feature (people)}
        \label{fig:dist_ori}
    \end{minipage}
    \hfill 
    \begin{minipage}[t]{0.49\textwidth}
        \centering
        \includegraphics[width=0.8\textwidth]{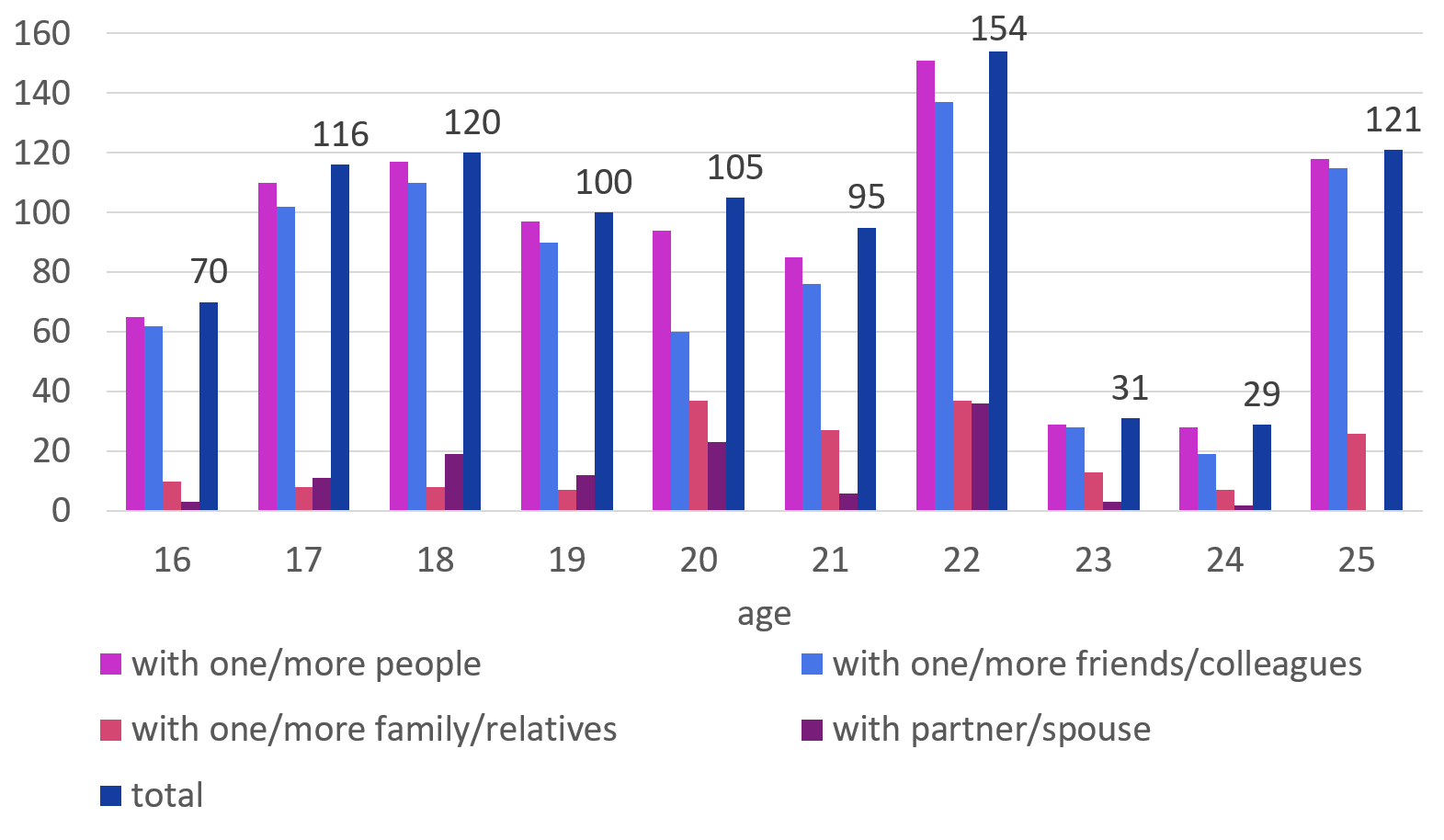}
        \caption{Self-Reports in terms of Age}
        \label{fig:dist_age}
    \end{minipage}
\end{center}
\vspace{-0.2 in}
\end{figure*}

\begin{figure*}[t]
\begin{center}
    
    \begin{minipage}[t]{0.49\textwidth}
        \centering
        \includegraphics[width=0.8\textwidth]{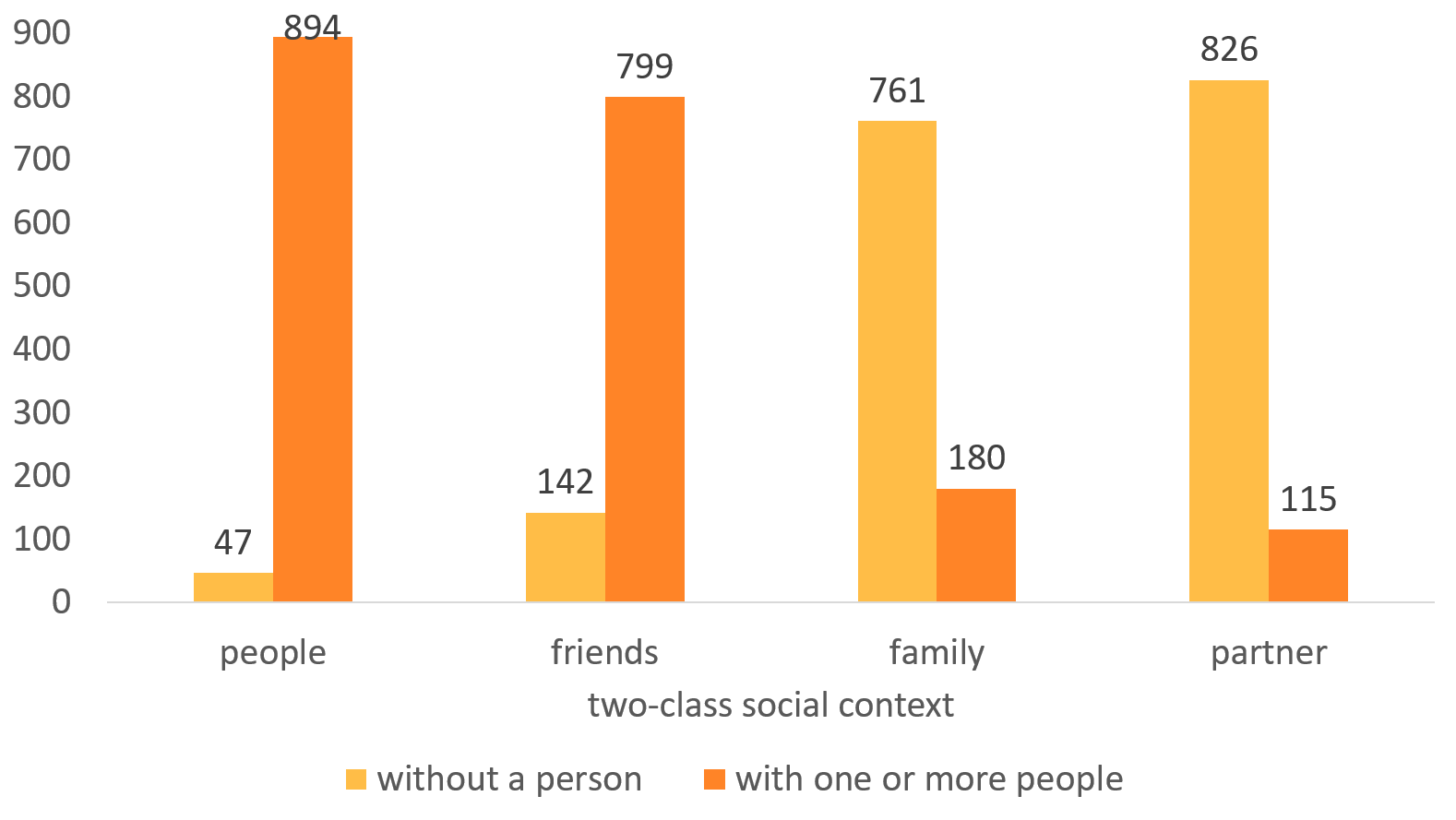}
        \caption{Distribution of two-class social contexts}
        \label{fig:dist_two_sc}
    \end{minipage}
    \hfill 
    \begin{minipage}[t]{0.49\textwidth}
        \centering
        \includegraphics[width=0.8\textwidth]{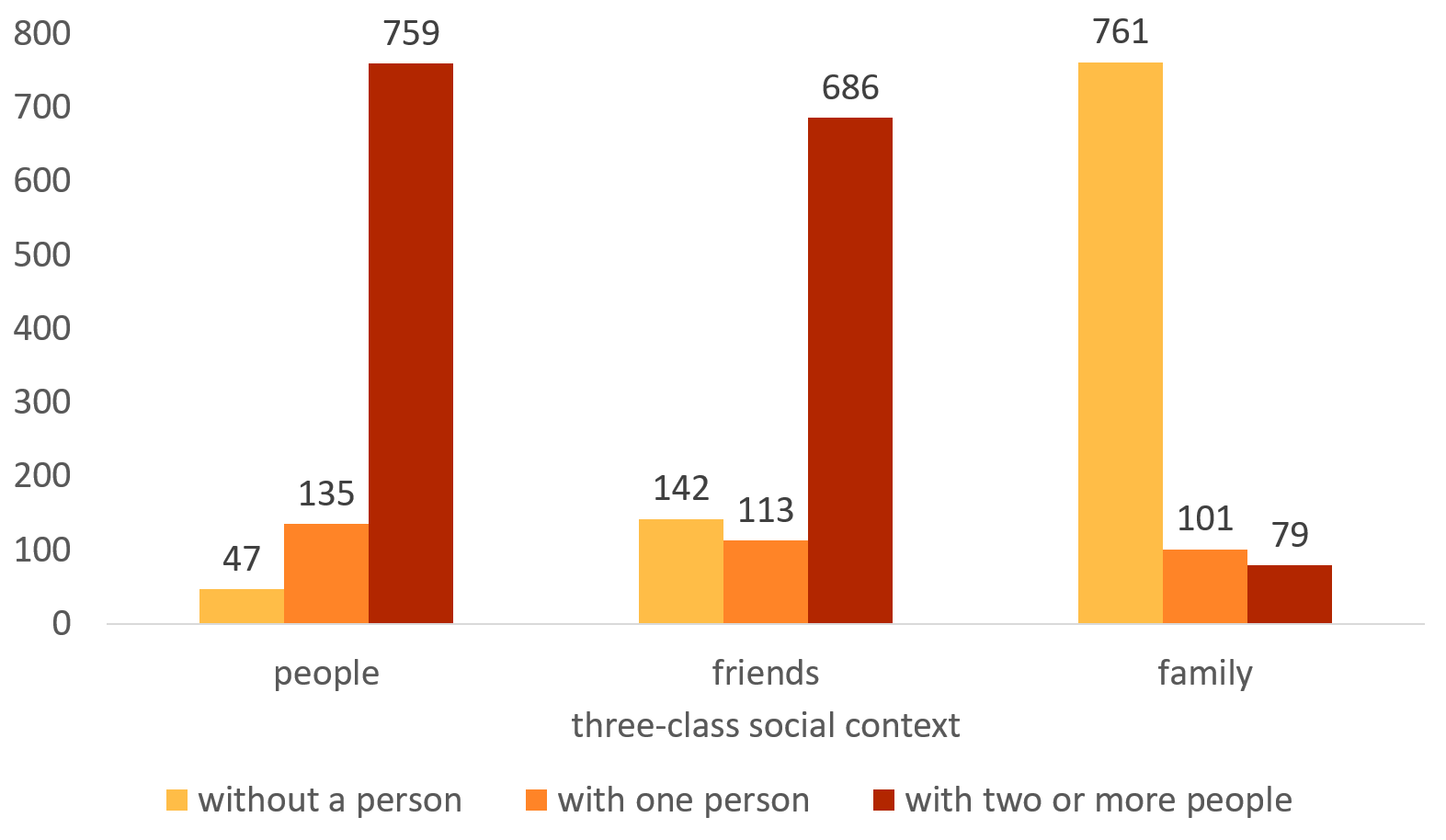}
        \caption{Distribution of three-class social contexts}
        \label{fig:dist_three_sc}
    \end{minipage}
\end{center}
\vspace{-0.2 in}
\end{figure*}

In this section, we provide a descriptive analysis regarding self-reports using demographic information, to understand the nature of the aggregate drinking behavior of participants.

\textbf{Self-Report Distribution for Different Social Contexts.} Figure~\ref{fig:dist_ori} provides a distribution of self-reports. Self-reports for partner is not shown here because it is only a binary response, and is included in Figure~\ref{fig:dist_two_sc}. Figure~\ref{fig:dist_two_sc} and Figure~\ref{fig:dist_three_sc} provide a distribution of self-reports for four two-class social contexts and three three-class social contexts, respectively. Results in Figure~\ref{fig:dist_two_sc} show that only 47 (5.0\%) drinking occasions were done alone as compared to 894 (95.0\%) occasions that were done with one/more people. Out of these 894 reports, 799 (89.4\%) were reported to have happened with two/more friends/colleagues. According to Figure~\ref{fig:dist_three_sc}, these 799 reports consist of 113 (14.1\%) reports that were done with one friend/colleague and 686 (85.9\%) reports that were done with two/more friends/colleagues, hence in a larger group. As a summary, participants consumed alcohol while being alone only on a small portion of occasions. This result is comparable to prior alcohol research, which shows that solitary drinking episodes are less frequent as compared to other social contexts \cite{Beck2008}. Moreover, the presence of two or more friends/colleagues was reported well over more than half of all drinking occasions (686/941 = 72.9\%). This result too is in line with prior work that state that young adults tend to drink alcohol for social facilitation and peer acceptance \cite{Beck2008}.

\begin{figure*}[t]
\begin{center}
    
    \begin{minipage}[t]{0.49\textwidth}
        \centering
        \includegraphics[width=0.8\textwidth]{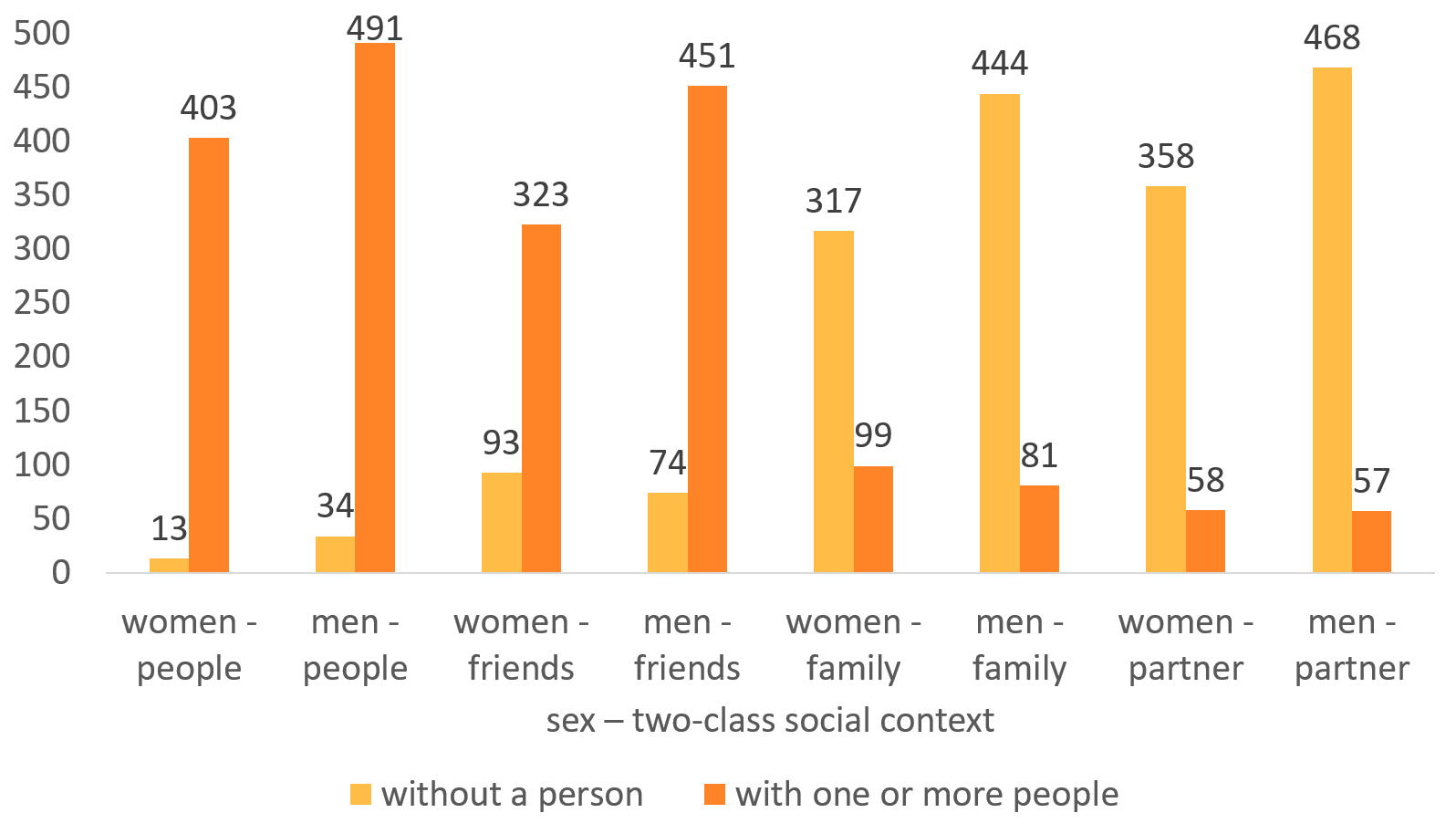}
        \caption{Self-Reports in terms of Sex and Two-Class Social Context}
        \label{fig:dist_sex_two}
    \end{minipage}
    \hfill 
    \begin{minipage}[t]{0.49\textwidth}
        \centering
        \includegraphics[width=0.8\textwidth]{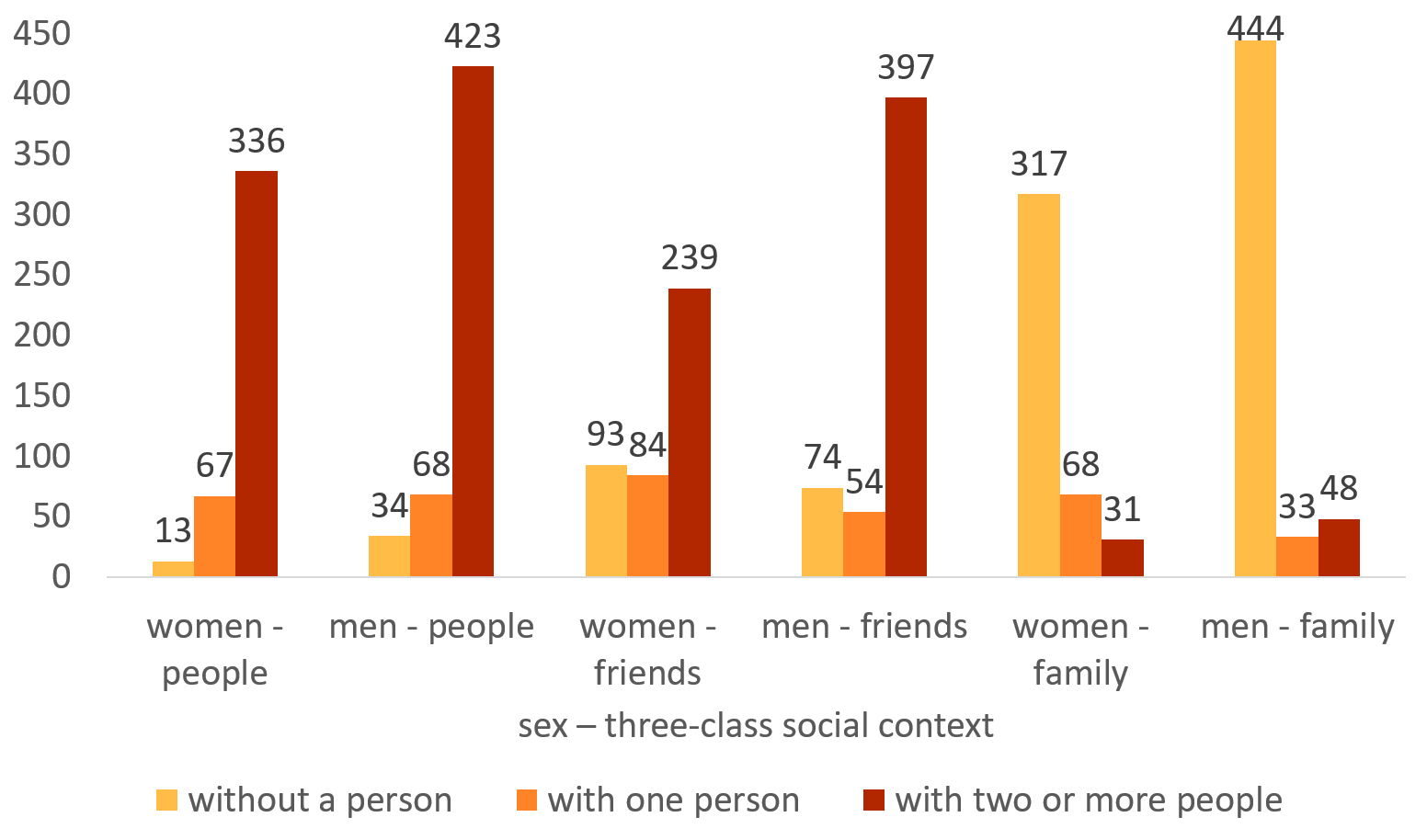}
        \caption{Self-Reports in terms of Sex and Three-Class Social Context}
        \label{fig:dist_sex_three}
    \end{minipage}
    
    \vspace{-0.2 in}
\end{center}
\end{figure*}

\textbf{Self-Report Distribution Breakdown Based on Sex.} In Figure~\ref{fig:dist_sex_two} and Figure~\ref{fig:dist_sex_three}, we present distributions of self-reports, based on sex and social context pairs. Results indicate that social contexts 'people' and 'friends' reported more drinking occasions with one/more people, for both men and women, whereas social contexts 'partner' and 'family' have significantly high number of drinking drinking events that were reported to be done alone. In addition, for the social context 'friends', Figure~\ref{fig:dist_sex_three} shows that the proportion of self-reports in groups of two/more (239) is just over half for women (239/416 = 57.5\%), whereas for men, drinking events with two/more friends/colleagues is 75.6\% (397/525), which is almost a 20\% difference for two sexes. This suggests that men reported a higher proportion of drinking occasions in groups of two/more people. This result is consistent with prior literature that state that men tend to drink in larger social contexts (specially with friends/colleagues) whereas women are less likely to do so \cite{thrul_associations_2018}. Further, women participants have reported drinking with family members 99 times (99/416 = 23.8\%), whereas men only reported to have done so 81 times (81/525 = 15.4\%), that is about 9\% less than women.

\textbf{Self-Report Distribution Breakdown Based on Age.} As shown in Figure~\ref{fig:dist_age}, participants' age ranged from 16 to 25. Except for ages 23 and 24 (31 and 29 self-reports, respectively), all other ages had over 70 self-reports. Moreover, the highest proportion of situations with one/more friends/colleagues (115/121 = 95.0\%) was reported by participants aged 25. The lowest proportion of situations with partner/spouse (0\%) was reported by the same age group. 
\section{Statistical Analysis (RQ1)}\label{sec:statistical_analysis}

\subsection{Pearson and Point-Biserial Correlation for Social Contexts and Passive Sensing Features}\label{subsec:corr1}

We conducted Pearson (PCC) \cite{Taylor1990} and Point-biserial (PBCC)  \cite{Bedrick2005} correlation analyses to measure the strength and the direction of the relationships between each of the three-class (takes values 0,1, and 2) and two-class (takes values 0 and 1) features and passive sensing features. The results of the top five features with the highest PCC or PBCC with each social context are summarized in Table~\ref{tab:correlation}. For a majority of social contexts (family$_{two}$, family$_{three}$, friends$_{two}$, friends$_{three}$, people$_{two}$, and people$_{three}$), multiple accelerometer-related features were among the top five features based on the correlation co-efficient values. The exception is the social context partner$_{two}$, where application features (e.g. food and drink) were among the top five. However, most of the values suggested, at best, weak positive or negative relationships.

\begin{table}[t]
    \centering
    \caption{Pearson Correlation Co-efficient (PCC) and Point-Biserial Correlation Co-efficient (PBCC) for Sensor Features and Social Contexts (two-class and three-class). With the top 5 features for each Social Context are included in the table. p-values are denoted with the following notation: p-value$\leq$10$^{-4}$:****; p-value$\leq$10$^{-3}$:***}.
    \resizebox{\textwidth}{!}{%
    \begin{tabular}{l l l l l l l l l }
    
    \rowcolor{red!5}
    &
    \multicolumn{2}{c}{\textbf{family}}&
    \multicolumn{2}{c}{\textbf{partner}}&
    \multicolumn{2}{c}{\textbf{friends}}&
    \multicolumn{2}{c}{\textbf{alone}}
    \\
    
    \rowcolor{red!5}
    &
    \textbf{Feature (Sensor)} & 
    \textbf{PBCC} &
    \textbf{Feature (Sensor)} &
    \textbf{PBCC} &
    \textbf{Feature (Sensor)} & 
    \textbf{PBCC} &
    \textbf{Feature (Sensor)} &
    \textbf{PBCC} 
    \\ \hline
    
    \multirow{5}{*}{\rotatebox[origin=c]{90}{\footnotesize{two-class}}}  &
    {angleXMax\_min (ACC)} & 
    0.19 (+) **** &  
    
    {yAxisAvgMax\_avg (ACC)} &  
    {0.14 (-)} **** &  
    
    mMean\_avg (ACC) &  
    0.22 (+) **** &  
    
    mMed\_max (ACC) & 
    {0.19} (+) ****  

    \\

    &
    {angleXMed\_min (ACC)} & 
    {0.19 (+)} ****&  
    
    {food\_and\_drink\_avg (APP)} &  
    {0.13 (+)} *** & 
    
    mSMAMean\_avg (ACC) &  
    0.22 (+) ****&  
    
    mSMAMed\_max (ACC) & 
    0.19 (+) ****  

    \\
    
    &
    {angleYAvg\_min (ACC)} & 
    {0.18 (+) ****} &  
    
    {food\_and\_drink\_min (APP)} &  
    {0.12 (+) ***} &  
    
    mMed\_max (ACC) &  
    0.22 (+) **** &  
    
    mMean\_max (ACC) & 
    0.19 (+) ****   

    \\

    &
    {angleXAvg\_min (ACC)} & 
    {0.18 (+) ****} &  
    
    {food\_and\_drink\_min (APP)} &  
    {0.12 (+) ***} &  
    
    mSMAMed\_max (ACC) &  
    0.22 (+) ****&  
    
    mSMAMean\_max (ACC) & 
    0.19 (+) ****  

    \\
    
    &
    {angleYMax\_min (ACC)} & 
    {0.18 (+) ****} &  
    
    {zAxisAvgMin\_avg (ACC)} &  
    {0.12 (+) ***} & 
    
    mMax\_avg (ACC) &  
    0.21 (+) ****&  
    
    dmListStd\_max (ACC) & 
    0.18 (+) ****  

    \\

    \rowcolor{red!5}
    &
    \textbf{Feature (Sensor)} & 
    \textbf{PCC} &
    -&
    -&
    \textbf{Feature (Sensor)} & 
    \textbf{PCC} &
    \textbf{Feature (Sensor)} &
    \textbf{PCC} 
    \\

    \multirow{5}{*}{\rotatebox[origin=c]{90}{\footnotesize{three-class}}}  &
    {angleXMed\_min (ACC)} & 
    0.18 (+) **** &  
    
    - &  
    - &  
    
    mMean\_avg (ACC) &  
    0.24 (+) **** &  
    
    mMean\_avg (ACC) & 
    {0.23} (+) **** 
    \\

    &
    {angleXMax\_min (ACC)} & 
    {0.18 (+) ****} &  
    
    - &  
    - & 
    
    mSMAMean\_avg (ACC) &  
    0.24 (+) **** &  
    
    mSMAMean\_avg (ACC) & 
    0.23 (+) ****   
    \\
    
    &
    {angleYAvg\_min (ACC)} & 
    {0.18 (+) ****} &  
    
    - &  
    - &  
    
    mMed\_avg (ACC) &  
    0.23 (+) **** &  
    
    yAxisAvgMin\_min (ACC) & 
    0.23 (-) ****   
    \\

    &
    {angleYMin\_min (ACC)} & 
    {0.17 (+) ****} &  
    
    - &  
    - &  
    
    mSMAMed\_avg (ACC) &  
    0.23 (+) ****&  
    
    mMean\_max (ACC) & 
    0.23 (+) ****  
    \\
    
    &
    {angleXMean\_min (ACC)} & 
    {0.17 (+) ****} &  
    
    - &  
    - & 
    
    mSMAMax\_avg (ACC) &  
    0.23 (+) ****&  
    
    mSMAMean\_max (ACC) & 
    0.23 (+) ****  
    \\ \hline
    
    \end{tabular}}
    \label{tab:correlation}
    
\end{table}

\subsection{Statistical Analysis of Dataset Features}\label{subsec:statistical_analysis}

\begin{table}[t]
    \centering
    \caption{t-statistic (T) (p-value$\leq$10$^{-4}$:****; p-value$\leq$10$^{-3}$:***; p-value$\leq$10$^{-2}$:**), and Cohen's-d (C) with 95\% confidence intervals (* if confidence interval include zero). Top five features are shown in decreasing order.}
    \resizebox{\textwidth}{!}{%
    \begin{tabular}{l l l l l l l l l l}

    \rowcolor{red!5}
    &
    \textbf{Feature} & 
    \textbf{T} &
    \textbf{Feature} &
    \textbf{C} &
    &
    \textbf{Feature} & 
    \textbf{T} &
    \textbf{Feature} &
    \textbf{C} 
    \\
    
    &
    \multicolumn{4}{c}{\textbf{family$_{three}$}}&
    &
    \multicolumn{4}{c}{\textbf{friends$_{three}$}}
    \\
    
    \cmidrule{2-5}
    \cmidrule{7-10}

    \multirow{5}{*}{\rotatebox[origin=c]{90}{alone vs. sgroup}}  &
    {blueStrengthMax\_avg (BLU)} & 
    {{4.41****}} &  
    {blueStrengthMed\_avg (BLU)} &  
    {0.42} &  
    \multirow{5}{*}{\rotatebox[origin=c]{90}{alone vs. sgroup}} &
    yAxisAvgMin\_max (ACC)& 
    4.05**** &  
    dmListMean\_max (ACC)&  
    0.37 
    \\

    &
     {blueStrengthAvg\_avg (BLU)} & 
     {4.21****} &  
     {zAxisAvgStd\_min (ACC)} &  
     {0.41} & 
     &
     xAxisAvgStd\_avg (ACC) &
    3.76*** &  
    xAxisAvgStd\_avg (ACC) &  
    0.34 
    \\
    
    &
     {blueStrengthMed\_avg (BLU)} & 
     {4.21****} &  
     {mMin\_min (ACC)} &  
     0.41 & 
     &
    anglexStd\_avg (ACC) &
   3.72*** &  
   anglexStd\_avg (ACC) &  
   0.33
    \\

    &
    {blueStrengthMin\_avg (BLU)} & 
    {3.87***} &  
    dmListStd\_min (ACC) &  
    0.40 &  
    &
    xAxisAvgStd\_max (ACC) & 
    3.65*** &  
   dmListMedian\_max (ACC) &  
   0.33 
    \\
    
    &
     {blueStrengthMax\_min (BLU)} & 
     {2.71**} &  
     {zAxisAvgMean\_min (ACC)} &  
     {0.40} & 
     &
     yAxisAvgMax\_avg (ACC) & 
    3.55*** &  
    angleyMin\_max (ACC) &  
    0.32
    \\
    \\

    \cmidrule{2-5}
    \cmidrule{7-10}

    \multirow{5}{*}{\rotatebox[origin=c]{90}{sgroup vs. lgroup}} &
    {zAxisAvgMean\_min (ACC)} & 
    3.22** &  
    {dmListMean\_min (ACC)} &  
    {0.38} &  
    \multirow{5}{*}{\rotatebox[origin=c]{90}{sgroup vs. lgroup}} &
    mMedian\_avg (ACC) & 
     3.44*** &  
    yAxisAvgMin\_avg (ACC) &  
     0.38 
  
    \\
    
    &
    zAxisAvgMedian\_min (ACC) & 
    3.08** &  
     {zAxisAvgStd\_min (ACC)} &  
     {0.36} &  
    &
     mMean\_avg (ACC) &
    3.43*** &  
    yAxisAvgMean\_avg (ACC) &  
    0.35
    
    \\
    
    &
    video\_players\_min (APP) & 
     2.84** &  
     {anglezMax\_min (ACC)} &  
     {0.34} &  
    &
   mSMAMedian\_avg (ACC) &
   3.41*** &  
   yAxisAvgMedian\_avg (ACC) &  
   0.35
    
    \\
    
    &
    zAxisAvgMax\_avg (ACC) & 
    2.71** &  
    system\_avg (APP) &  
    {0.32} &  
     &
    mSMAMean\_avg (ACC) & 
    3.41*** &  
   mMax\_min (ACC) &  
   0.33 
    \\
    
    &
    video\_players\_avg (APP) & 
     2.70** &  
     zAxisAvgMax\_avg (ACC) &  
      0.32 &  
     &
     mNewStd\_avg (ACC) & 
    3.15*** &  
    mMedian\_avg (ACC) &  
    0.33
    \\
    \\

    \cmidrule{2-5}
    \cmidrule{7-10}
    
    \multirow{5}{*}{\rotatebox[origin=c]{90}{alone vs. lgroup}} &
    system\_min (APP) & 
    5.83**** &  
    mMin\_min (ACC) &  
    0.49 &  
    \multirow{5}{*}{\rotatebox[origin=c]{90}{alone vs. lgroup}} &
    mMean\_avg (ACC) & 
    8.61**** &  
    mMean\_avg (ACC) &  
    0.56  
    \\
    
    &
    anglexMax\_min (ACC) & 
    5.64**** &  
    zAxisAvgStd\_min (ACC) &  
    0.49 &  
    &
    mSMAMean\_avg (ACC)&
    8.60**** &  
    mSMAMean\_avg (ACC)&  
    0.56 
    
    \\
    
    &
    anglexMedian\_min (ACC) & 
    5.64**** &  
    anglezMedian\_min (ACC) &  
    0.48 &  
    &
    mMedian\_max (ACC) &
    8.44**** &  
    mMedian\_max (ACC) &  
    0.55
    
    \\
    
    &
    angleyMean\_min (ACC) & 
    5.55**** &  
    anglexMax\_min (ACC) &  
    0.47 &  
    &
    mSMAMedian\_max (ACC) & 
    8.40**** &  
    mSMAMedian\_avg (ACC) &  
    0.55  
    \\
    
    &
    angleyMin\_min (ACC) & 
    5.54**** &  
    zAxisAvgMean\_min (ACC) &  
    0.46 &  
    &
    mMax\_avg (ACC) & 
    8.40**** &  
    mMax\_avg (ACC) &  
    0.54  
    \\ 
    \\ \hline
    
    &
    \multicolumn{4}{c}{\textbf{people$_{three}$}}&
    &
    \multicolumn{4}{c}{\textbf{people$_{two}$}}
    \\

    \cmidrule{2-5}
    \cmidrule{7-10}

    \multirow{5}{*}{\rotatebox[origin=c]{90}{alone vs. sgroup}}  &
    mSMAMedian\_max (ACC) & 
    3.81*** &  
    zAxisAvgMedian\_max (ACC) &  
    0.41 &  
    \multirow{5}{*}{\rotatebox[origin=c]{90}{alone vs. group}} &
    yAxisAvgMin\_max (ACC) & 
    4.05**** &  
    dmListMean\_max (ACC) &  
    0.37 
    \\

    &
    mMedian\_max (ACC) & 
     3.81*** &  
   xAxisAvgStd\_max (ACC) &  
     0.41 & 
     &
     xAxisAvgStd\_avg (ACC) &
    3.76*** &  
    xAxisAvgStd\_avg (ACC) &  
    0.34 
    \\
    
    &
    anglexStd\_max (ACC) & 
    3.79*** &  
    xAxisAvgMedian\_min (ACC) &  
    0.40 & 
     &
    anglexStd\_avg (ACC) &
   3.72*** &  
   anglexStd\_avg (ACC) &  
   0.33
    \\
    
    &
    xAxisAvgStd\_max (ACC) & 
    3.63*** &  
    xAxisAvgMean\_max (ACC) &  
    0.40 &  
    &
    xAxisAvgStd\_max (ACC) & 
    3.65*** &  
   dmListMedian\_max (ACC) &  
   0.33 
    \\
    
    &
    mMean\_max (ACC) & 
    3.60*** &  
    angleyMax\_max (ACC) &  
    0.38 & 
     &
     yAxisAvgMax\_avg (ACC) & 
    3.55*** &  
    angleyMin\_max (ACC) &  
    0.32  
    \\
    
    \cmidrule{2-5}
    \cmidrule{7-10}

    &
     & 
     &  
     &  
     &  
    &
    \multicolumn{4}{c}{\textbf{family$_{two}$}}
    \\
    
    \cmidrule{7-10}
    
    \multirow{5}{*}{\rotatebox[origin=c]{90}{sgroup vs. lgroup}} &
    yAxisAvgMin\_min (ACC) & 
    5.30**** &  
    mNewStd\_avg (ACC) &  
    0.43 &  
    \multirow{5}{*}{\rotatebox[origin=c]{90}{alone vs. group}} &
    anglexMax\_min (ACC)& 
    6.82**** &  
    mMin\_min (ACC)&  
    0.46 
    \\
    
    &
    yAxisAvgMin\_avg (ACC)& 
    5.11**** &  
    mMean\_avg (ACC)&  
    0.42 &  
    &
   anglexMedian\_min (ACC)& 
    6.82**** &  
    zAxisAvgStd\_min (ACC) &  
    0.46 
    \\
    
    &
    yAxisAvgMean\_min (ACC)& 
    5.10**** &  
    mSMAMean\_avg (ACC)&  
    0.42 &  
     &
    angleyMean\_min (ACC)& 
    6.64**** &  
    anglezMedian\_min (ACC)&  
    0.44 
    \\
    
    &
    yAxisAvgMedian\_min (ACC)& 
    5.07**** &  
    mSMAMax\_avg (ACC)&  
    0.42 &  
     &
    anglexMean\_min (ACC)& 
    6.49**** &  
    dmListStd\_min (ACC)&  
    0.43 
    \\
    
    &
    yAxisAvgMax\_min (ACC)& 
    4.18**** &  
    mMax\_avg (ACC)&  
    0.41 &  
    &
    angleyMax\_min (ACC)& 
     6.42**** &  
    zAxisAvgMean\_min (ACC)&  
    0.43 
    \\
    
    \cmidrule{2-5}
    \cmidrule{7-10}
    
    &
     & 
     &  
     &  
     &  
    &
    \multicolumn{4}{c}{\textbf{partner$_{two}$}}
    \\
    
    \cmidrule{7-10}
    
    \multirow{5}{*}{\rotatebox[origin=c]{90}{alone vs. lgroup}} &
    yAxisAvgMin\_min (ACC)& 
    7.44**** &  
    xAxisAvgStd\_max (ACC)&  
    0.77 &  
    \multirow{5}{*}{\rotatebox[origin=c]{90}{alone vs. group}} &
    food\_and\_drink (APP)& 
    4.55**** &  
    yAxisAvgMax\_avg (ACC)&  
    0.43 
    \\
    
    &
    zAxisAvgMin\_min (ACC) & 
    6.76**** &  
    zAxisAvgMedian\_max (ACC) &  
    0.77 &  
    &
    food\_and\_drink (APP) & 
    4.54**** &  
    zAxisAvgMin\_avg (ACC) &  
    0.33 
    \\
    
    &
    xAxisAvgMin\_min (ACC) & 
    6.63**** &  
    angleyMax\_max (ACC) &  
    0.75 &  
    &
    food\_and\_drink (APP) & 
    4.55**** &  
    proximityRecord\_avg (PRO) &  
    0.29 
    \\
    
    &
    yAxisAvgMedian\_min (ACC)& 
    6.54**** &  
    anglezMin\_max (ACC)&  
    0.75 &  
     &
    zAxisAvgMin\_avg (ACC)& 
    4.17**** &  
    yAxisAvgStd\_avg (ACC)&  
    0.27 
    \\
    
    &
    yAxisAvgMean\_min (ACC) & 
    6.31**** &  
    mMedian\_avg (ACC) &  
    0.73 &  
    &
    zAxisAvgMean\_avg (ACC) & 
    3.35**** &  
    angleyStd\_avg (ACC) &  
    0.27 
    \\

    \cmidrule{2-5}
    \cmidrule{7-10}
    
    &
    &
    &
    &
    &
    &
    \multicolumn{4}{c}{\textbf{friends$_{two}$}}
    \\

    \cmidrule{7-10}

     &
    - & 
    - &  
    - &  
    - &  
    \multirow{5}{*}{\rotatebox[origin=c]{90}{alone vs. group}} &
    mMean\_avg (ACC) & 
    8.10**** &  
    mMean\_avg (ACC) &  
    0.52 
    \\

    &
    - & 
    - &  
    - &  
    - &  
     &
    mSMAMean\_avg (ACC) &
    8.08*** &  
    mSMAMean\_avg (ACC) &  
    0.52
    \\
    
    &
    - & 
    - &  
    - &  
    - &  
     &
    mMedian\_max (ACC) &
   7.97**** &  
   mMedian\_avg (ACC) &  
   0.50
    \\
    
    &
    - & 
    - &  
    - &  
    - &  
    &
   mSMAMedian\_max (ACC) & 
   7.94**** &  
   mSMAMedian\_avg (ACC) &  
   0.50 
    \\
    
    &
    - &
    - &
    - &
    - &
    &
    mMax\_avg (ACC) &
    7.89**** &
    yAxisAvgStd\_max (ACC) &
    0.50 \\ \\ \hline

    \end{tabular}
    }
    \label{tab:tstatistic}
    \vspace{-0.2 in}
\end{table}

Table~\ref{tab:tstatistic} shows statistics such as t-statistic \cite{Kim2015}, p-value \cite{Greenland2016}, and Cohen's-d (effect size) with 95\% confidence interval (CI) \cite{Lakens2013} for the top five features in the dataset for the seven different social contexts. For two-class social contexts, the objective is to identify passive sensing features that help discriminate between: without people (alone) and with one/more people (group). Here, the term group is used because it could either be a small group of two to three people, or a large group of more than ten people. Further, for three-class social contexts, the objective is to identify passive sensing features that help discriminate between: (a) without people (alone) vs. with one person (sgroup); (b) with one person (sgroup) vs. with two/more people (lgroup); and (c) without people (alone) vs. with two/more people (lgroup), where sgroup and lgroup stands for small group and large group, respectively. The features are ordered by the descending order of t-statistics and Cohen's-d values. In addition, prior work stated the lack of sufficient informativeness in p-values \cite{Yatani2016, Lee2016}. For this reason, we calculated the Cohen's-d \cite{Rice2005} to measure the statistical significance of features. We adopted the following rule-of-thumb, commonly used to interpret Cohen's-d values: 0.2 = small effect size; 0.5 = medium effect size; and 0.8 = large effect size. According to this notion, the higher the value of Cohen's-d, the higher the possibility of discerning the two groups using the feature. In addition, 95\% confidence intervals for Cohen's-d were calculated, and if the interval does not overlap with zero, the difference can be considered as significant \cite{Lee2016}.

For the social context family$_{three}$, features from the bluetooth sensor were among the top five in terms of t-statistic and Cohen's-d, for the combination alone vs. sgroup. In addition, all the top five features had Cohen's-d values closer to medium effect size. Further, a total of 122 features had Cohen's-d values above small effect size and confidence intervals not including zero. For the combinations sgroup vs. lgroup and alone vs. lgroup, the majority of features were from the accelerometer and two features (video\_player and system) were from application usage. In addition, if the hierarchy of the social contexts alone, sgroup, and lgroup is considered, sgroup is in the middle, sandwiched by alone and lgroup, that are further apart, hence, it would be easier to discern between these two groups. This is indicated in the results for the combination alone vs. lgroup, that have higher t-statistics and Cohen's-d values (some around medium effect size) compared to the other two combinations (alone vs. sgroup and sgroup vs. lgroup). Furthermore, for the social contexts friends$_{three}$ and people$_{three}$, for all three combinations, all features in the top five in terms of both t-statistic and Cohen's-d are from the accelerometer. In addition, for friends$_{three}$, features in the combination alone vs. lgroup had high t-statistics and Cohen's-d values above medium effect size. In fact, 14 features, all of which are from the accelerometer had Cohen's-d values above medium effect size. In addition, for people$_{three}$, 44 features had Cohen's-d values above medium effect size, and the highest ones were closer to large effect size, meaning that these accelerometer feature could discriminate between alone and lgroup social contexts. 

For two-class social contexts people$_{two}$, family$_{two}$, and friends$_{two}$, all features in the top five for both t-statistic and Cohen's-d were from the accelerometer. Further, only friends$_{two}$ had features with Cohen's-d above medium effect size among all four two-class social contexts. However, for partner$_{two}$, several features from application usage (food and drink app usage) were among the top five for t-statistics. In addition, a feature from the proximity sensor had a Cohen's-d of 0.29, which is above small effect size. As a summary, results from the statistical analysis suggest that for all the social contexts, accelerometer features could be informative of the group dynamic. In addition, for social contexts related to partner/spouse, app usage behavior and proximity sensors could be informative. Moreover, bluetooth sensor had high statistical significance in discriminating social contexts related to family members. 

\section{Social Context Inference}\label{sec:inference}

\subsection{Two-Class and Three-Class Social Context Inference (RQ2)}\label{subsec:inference}

In this section, we use all the available smartphone sensing features and implement seven social context tasks, using features defined in Section~\ref{sub:social_context} as target variables. The tasks include four two-class inference tasks and three three-class inference tasks: (1) family$_{two}$, (2) partner$_{two}$, (3) friends$_{two}$, (4) people$_{two}$, (5) family$_{three}$, (6) friends$_{three}$, and (7) people$_{three}$. In this phase, we used scikitlearn \cite{scikit-learn} and keras \cite{chollet2015} frameworks together with python, and conducted experiments with several model types: (1) Random Forest Classifier \cite{Cutler2011}, (2) Naive Bayes \cite{Rish2001}, (3) Gradient Boosting \cite{Natekin2013}, (4) XGBoost \cite{Chen2016}, and (5) AdaBoost \cite{Schapire2013}. These models were chosen by considering the tabular nature of the dataset, interpretability of results, and small size of the dataset. In addition, we used the leave k-participants out strategy (k = 20) when conducting experiments, where testing and training splits did not have data from the same user, hence avoiding possible biases in experiments. Further, similar to recent ubicomp studies \cite{Bae2018, Meegahapola2021v2, Kondo2019}, we used the Synthetic Minority Over-sampling Technique (SMOTE) \cite{Chawla2002} to obtain training sets for each inference task. As recommended by Chawla et al. \cite{Chawla2002}, when and where necessary, we under-sampled the majority class/classes to match over-sampled minority class/classes to create balanced datasets, hence not over-sampling unnecessarily beyond doubling the minority class size. In addition, we also calculated the area under the curve (AUC) (for three-class inferences, one vs. the rest technique, using macro averaging) using the receiver operator characteristics (ROC) curves. All experiments were repeated for ten iterations. We report mean and standard deviation of accuracies, and mean of AUC using results from the ten iterations.

Table~\ref{tab:inference_results} summarizes the results of the experiments. All the two-class inference tasks achieved accuracies over 80\%. Moreover, all the three-class inferences achieved accuracies over 75\%. When considering model types, Random Forest classifiers performed the best across five out of the seven inference tasks (family$_{two}$, partner$_{two}$, friends$_{two}$, family$_{three}$, and people$_{three}$) and Gradient Boosting had higher accuracies for two inference tasks (people$_{two}$ and friends$_{three}$). Generally, all models included in the study, except for Naive Bayes, performed reasonably well. Further, low standard deviation values suggest that regardless of the samples used for training and testing, the models generalized reasonably well. AUC scores followed a similar trend as the accuracy. These results suggest that passive mobile sensing features could be used to infer both two-class and three-class social contexts related to alcohol consumption, with reasonable performance.

\begin{table}[t]
        \small
        \centering
        \caption{Mean (\={A}) and Standard Deviation (A$_{\sigma}$) of inference accuracies and the mean area under the curve of the receiver operator characteristic curve (AUC), calculated from 10 iterations, using five different models, for two-class and three-class tasks, with attributes such as family, friends/colleagues, spouse/partner, and alone. Results are presented as: \={A} (A$_{\sigma}$), AUC}
        \resizebox{0.77\textwidth}{!}{%
        \begin{tabular}{c l l l l l l}

        \rowcolor{red!5}
        &
        \textbf{Target Variable}& 
        \textbf{Random Forest} &
        \textbf{XG Boost} &
        \textbf{Ada Boost} &
        \textbf{Gradient Boost}&
        \textbf{Naive Bayes} 
        \\ \hline 
        
        \multirow{4}{*}{\rotatebox[origin=c]{90}{\footnotesize{\textbf{two-class}}}}&
        baseline &
        50.0 (0.0), 50.0 & 
        50.0 (0.0), 50.0 & 
        50.0 (0.0), 50.0 & 
        50.0 (0.0), 50.0 & 
        50.0 (0.0), 50.0 
        \\

        &
        family$_{two}$  &
        \textbf{86.1} (3.1), 84.8 & 
        82.6 (3.4), 73.2 & 
        82.5 (4.2), 71.7& 
        83.2 (3.7), 74.2 & 
        65.6 (6.9), 67.6 
        \\
        
        &
        partner$_{two}$  &
        \textbf{87.4} (2.6), 82.6 & 
        84.6 (4.6), 74.7 & 
        83.5 (3.8), 69.3 & 
        84.7 (5.1), 78.8 & 
        68.6 (8.2), 64.2 
        \\
        
        &
        friends$_{two}$  &
        \textbf{80.1} (2.9), 81.3 & 
        78.3 (4.5), 75.2 & 
        78.0 (4.1), 70.4 & 
        78.7 (3.7), 73.7 & 
        64.0 (4.2), 61.4 
        \\

        &
        people$_{two}$ &
        83.3 (3.2), 79.2 & 
        84.1 (3.1), 75.2 & 
        82.7 (4.3), 79.5 & 
        \textbf{84.2} (2.8), 76.9 & 
        72.3 (4.7), 67.3  
        \\
        
        \cmidrule{2-7}
        
        \multirow{4}{*}{\rotatebox[origin=c]{90}{\footnotesize{\textbf{three-class}}}}&
        baseline &
        33.3 (0.0), 50.0 & 
        33.3 (0.0), 50.0 & 
        33.3 (0.0), 50.0 & 
        33.3 (0.0), 50.0 & 
        33.3 (0.0), 50.0 
        \\

        &
        family$_{three}$  &
        \textbf{85.9} (2.2), 80.5 & 
        81.4 (3.1), 73.2 & 
        73.9 (3.7), 63.2 & 
        83.0 (2.6), 70.2 & 
        63.1 (4.2), 61.8 
        \\

        &
        friends$_{three}$  &
        76.7 (2.3), 78.2 & 
        77.1 (3.6), 70.1 & 
        71.0 (3.1), 68.8 & 
        \textbf{77.6} (2.8), 72.3 & 
        62.2 (5.1), 63.5 
        \\
        
        &
        people$_{three}$  &
        \textbf{78.3} (2.1), 75.3 & 
        71.2 (2.6), 69.8 & 
        67.1 (3.8), 67.8 & 
        73.8 (3.2), 73.1 & 
        57.9 (6.7), 67.2 \\ \hline 
        
        \end{tabular}}
        \label{tab:inference_results}

\end{table}

\subsection{Social Context Inference for Different Sensors (RQ2)}\label{subsec:diff_sensors_inference}

Prior work in mobile sensing has argued for multiple inference models for the same inference task, in the case of sensor failure \cite{Yao2018, Meegahapola2021, Santani2018}. For instance, during a weekend night, young adults could be concerned for the battery life of their phone, and could turn-off bluetooth, wifi, and location sensors that drain the battery faster. In such cases, having separate inference models that use different data sources to infer the same target attribute could be beneficial. In addition, prior work has segregated passive sensing modalities into Continuous Sensing (using embedded sensors in the smartphone) and Interaction Sensing (sensing the users' phone usage and interaction behavior) \cite{Meegahapola2021}. Considering these aspects, we conducted experiments for different feature groups based on the sensing modality (accelerometer, applications, battery, bluetooth, proximity, location, screen, and wifi) and the following feature group combinations that are meaningful in the context of drinking and young adults:

\begin{itemize}
    \item \textbf{Continuous Sensing (ConSen):} These sensing modalities use embedded sensors to capture context. Examples are accelerometer, battery, bluetooth, proximity, location, and wifi. ConSen contains features from all these sensing modalities, and this feature group combination can measure the capability of the smartphone in inferring the social context of drinking, even if the user does not necessarily use the smartphone, because the considered sensing modalities sample data regardless of the phone usage behavior.  
    \item \textbf{Interaction Sensing (IntSen):} These sensing modalities capture the phone usage and interaction behavior. Examples include screen events and application usage. In addition, these sensing modalities do not fail often because there is no straightforward way for users to turn-off interaction sensing modalities. Furthermore, these sensing modalities consume far less power compared to continuous sensing. In this context, this feature group combination could measure the capability of a smartphone to infer the social context of drinking, based on the way young adults use and interact with the smartphone. 
    
\end{itemize}{}

For the two above mentioned feature groups, we conducted experiments using the same procedure as given in Section~\ref{subsec:diff_sensors_inference}. Even though we got results for all models, we only present results for random forest classifiers in Table~\ref{tab:inference_results_v2} because they output feature importance values which are useful to interpret results in Section~\ref{subsec:feature_importance}, and they provide the results with highest accuracy and AUC values for a majority of inference tasks. Even though the accuracies were well above baselines for both two-class and three-class inference tasks, the lowest accuracies were recorded for SCR. This could either be because of the far too small dimensionality (only three features) or because the features were less informative. For the inference of social context partner$_{two}$, APP provided the highest accuracy of 82.92\% followed by ACC that provided an accuracy of 81.21\%. This suggests that the app usage behavior during drinking events is informative of whether participants are with a partner/spouse or not. This could also be related to prior work regarding \textit{partner phubbing} \cite{Roberts2016, Chotpitayasunondh2016} that could lead to relationship dissatisfaction and disappointment. People might try to avoid phubbing (hence use the phone less/differently than normal) when they are with their partner/spouse. Furthermore, except for this inference, for all other social context inferences, the highest accuracies were obtained using ACC (in the range of 71.52\% to 83.33\%). This suggests that physical activity levels and movement dynamics around drinking events could be used to infer social contexts such as family (family$_{two}$ and family$_{three}$), friends (family$_{two}$ and family$_{three}$), and people (people$_{two}$ and people$_{three}$). In addition, results from the AUC followed a similar trend to accuracies. For two-class inferences, except for SCR, all other modalities reported AUC scores above 70\%. However, for three-class inferences, except for ACC, all other modalities reported AUC scores below 70\%. Further, except for SCR, for all the other inferences, standard deviation scores were reasonably low, suggesting that inference results hold regardless of the training and testing splits. High standard deviations for SCR could be because of the low number of features, which was also reflected with low accuracies and AUC scores.

Feature group combinations ConSen and IntSen provided similar accuracies for all inference tasks even though ConSen achieved slightly better than IntSen for each inference. While ConSen outperformed ACC and APP for all the inferences, IntSen had slightly lower accuracies for social contexts family$_{three}$ (82.36\%) and friend$_{three}$ (71.44\%) as compared to ACC, that had accuracies 82.60\% and 71.52\% for the respective inferences. Standard deviation scores for both ConSen and IntSen were low. In addition, AUC scores too were above 70\% for all cases, which is a reasonable result. Finally, the results suggest that IntSen could provide reasonably high accuracies as compared to ConSen in case of sensor failure, and in the worst case scenario, ACC provides fair accuracies for all the inference tasks, which is satisfactory given that it is just one sensing modality.    

\begin{table}[t]
        \small
        \centering
        \caption{Social Context Inference accuracy breakdown for sensor type based feature groups and feature group combinations using Random Forest classifiers. Both the mean (\={A}) and standard deviation (A$_{\sigma}$) of accuracies from cross validation are reported in addition to the mean area under the curve (AUC) from reciever operator characteristics graph (ROC)}
        \resizebox{\textwidth}{!}{%
        \begin{tabular}{l l l l l l l l l l l l l l l l }

        \rowcolor{red!5}
        \textbf{Feature Group} & 
        \multicolumn{8}{c}{\textbf{two-class}} &
        &
        \multicolumn{6}{c}{\textbf{three-class}} 
        \\
        
        \arrayrulecolor{red!5} \specialrule{6pt}{0pt}{-6pt} \arrayrulecolor{black}
        \cmidrule{2-9}
        \cmidrule{11-16}
        
        \rowcolor{red!5}
        \textbf{(\# of features)}& 
        \textbf{family$_{two}$} & 
        &
        \textbf{partner$_{two}$} & 
        &
        \textbf{friends$_{two}$} & 
        &
        \textbf{people$_{two}$}& 
        &
        &
        \textbf{family$_{three}$} & &
        \textbf{friends$_{three}$} & &
        \textbf{people$_{three}$} &
        
        \\
        \rowcolor{red!5}
        &
        \={A} (A$_{\sigma}$), AUC & 
        &
        \={A} (A$_{\sigma}$), AUC & 
        &
        \={A} (A$_{\sigma}$), AUC & 
        &
        \={A} (A$_{\sigma}$), AUC & 
        &
        &
        \={A} (A$_{\sigma}$), AUC & 
        &
        \={A} (A$_{\sigma}$), AUC & 
        &
        \={A} (A$_{\sigma}$), AUC & 
        
        \\ \hline 
        
        Baseline &
        50.0 (0.0), 50.0& 
        &
        50.0 (0.0), 50.0& 
        &
        50.0 (0.0), 50.0& 
        &
        50.0 (0.0), 50.0& 
        &
        &
        33.3 (0.0), 50.0& 
        &
        33.3 (0.0), 50.0& 
        &
        33.3 (0.0), 50.0&  
        
        \\

        ACC (150) &
        83.3 (2.4), 80.2& 
        &
        81.2 (3.1), 79.6 & 
        &
        74.9 (3.0), 72.5 & 
        &
        81.1 (3.1), 81.4 & 
        &
        &
        82.6 (1.9), 78.5 & 
        &
        71.5 (2.5), 70.1 & 
        &
        72.4 (2.7), 72.0 &  
        
        \\

        \rowcolor{gray!5}
        APP (105)&
        82.9 (3.5), 80.7 & 
        &
        82.9 (3.0), 79.2 & 
        &
        74.3 (2.4), 76.7 & 
        &
        80.5 (2.5), 81.1 & 
        &
        &
        81.4 (2.4), 81.5 & 
        &
        69.5 (3.1), 69.1 & 
        &
        71.9 (2.3), 70.2 &   
    
        \\

        BAT (36)&
        78.7 (2.8), 76.7 & 
        &
        77.5 (3.6), 77.3 & 
        &
        71.5 (3.0), 73.6 & 
        &
        78.1 (3.3), 72.1 & 
        &
        &
        77.5 (2.7), 75.6 & 
        &
        66.1 (3.1), 67.8 & 
        &
        68.1 (2.6), 68.4 &   

        \\
        
        \rowcolor{gray!5}
        BLU (27)&
        74.6 (2.8), 72.1 & 
        &
        75.5 (3.6), 73.8 & 
        &
        69.0 (2.8), 70.3 & 
        &
        74.0 (3.1), 73.1 & 
        &
        &
        69.3 (2.9), 68.8 & 
        &
        59.4 (2.7), 61.4 & 
        &
        60.5 (2.8), 66.4 &   
        
        \\

        PRO (18)&
        74.1 (3.1), 71.9 & 
        &
        75.6 (2.3), 74.5 & 
        &
        69.9 (2.8), 68.2 & 
        &
        75.8 (2.7), 76.8 & 
        &
        &
        70.4 (2.2), 73.2 & 
        &
        59.1 (3.8), 61.9 & 
        &
        60.7 (2.4), 62.9 &   
        
        \\

        \rowcolor{gray!5}
        LOC (39)&
        79.2 (3.0), 77.1 & 
        &
        78.9 (2.7), 78.2 & 
        &
        74.1 (3.2), 76.1 & 
        &
        77.6 (2.6), 76.9 & 
        &
        &
        77.2 (2.6), 76.4 & 
        &
        67.0 (3.3), 69.1 & 
        &
        69.5 (2.9), 68.7 &   
        
        \\

        SCR (3) &
        68.3 (4.5), 61.1 & 
        &
        69.7 (4.7), 60.3 & 
        &
        64.5 (4.6), 62.8 & 
        &
        71.9 (3.1), 67.2 & 
        &
        &
        62.2 (5.6), 60.7 & 
        &
        54.1 (4.3), 55.2 & 
        &
        54.8 (5.5), 56.1 &  
        
        \\

        \rowcolor{gray!5}
        WIF (36)&
        77.5 (2.9), 78.1 & 
        &
        77.1 (2.9), 76.9 & 
        &
        68.8 (3.7), 70.2 & 
        &
        75.3 (3.9), 76.1 & 
        &
        &
        73.1 (1.9), 73.0 & 
        &
        61.6 (2.9), 63.6 & 
        &
        64.5 (2.8), 67.1 &   
        
        \\

        \cmidrule{2-9}
        \cmidrule{11-16}
        
        ConSen (306) &
        85.7 (2.6), 82.1 & 
        &
        86.8 (2.9), 80.8 & 
        &
        79.5 (3.2), 80.2 & 
        &
        82.9 (2.1), 81.4 & 
        &
        &
        85.3 (1.9), 76.8 & 
        &
        76.7 (2.7), 76.9 & 
        &
        77.9 (3.0), 76.1 &   
        \\

        \rowcolor{gray!5}
        IntSen (96) &
        83.3 (2.0), 80.1 & 
        &
        83.1 (2.5), 81.7 & 
        &
        76.5 (2.9), 76.3 & 
        &
        81.6 (2.5), 81.5 & 
        &
        &
        82.3 (2.7), 78.2 & 
        &
        71.4 (2.7), 71.2 & 
        &
        73.2 (2.7), 72.7 &   
        
        \\

        \cmidrule{2-9}
        \cmidrule{11-16}
        
        ALL (402) &
        86.1 (3.1), 84.8 & 
        &
        87.4 (2.6), 82.6& 
        &
        80.1 (2.9), 81.3 & 
        &
        83.3 (3.2), 79.2 & 
        &
        &
        85.9 (2.2), 80.5 & 
        &
        76.7 (2.3), 78.2 & 
        &
        78.3 (2.1), 75.3 &   
        
        \\ \hline
        
        \end{tabular}}
        \label{tab:inference_results_v2}
\vspace{-0.2 in}
\end{table}

\begin{figure*}[t]
\begin{center}
    \begin{subfigure}[t]{0.33\textwidth}
        \centering
        \includegraphics[width=0.78\textwidth]{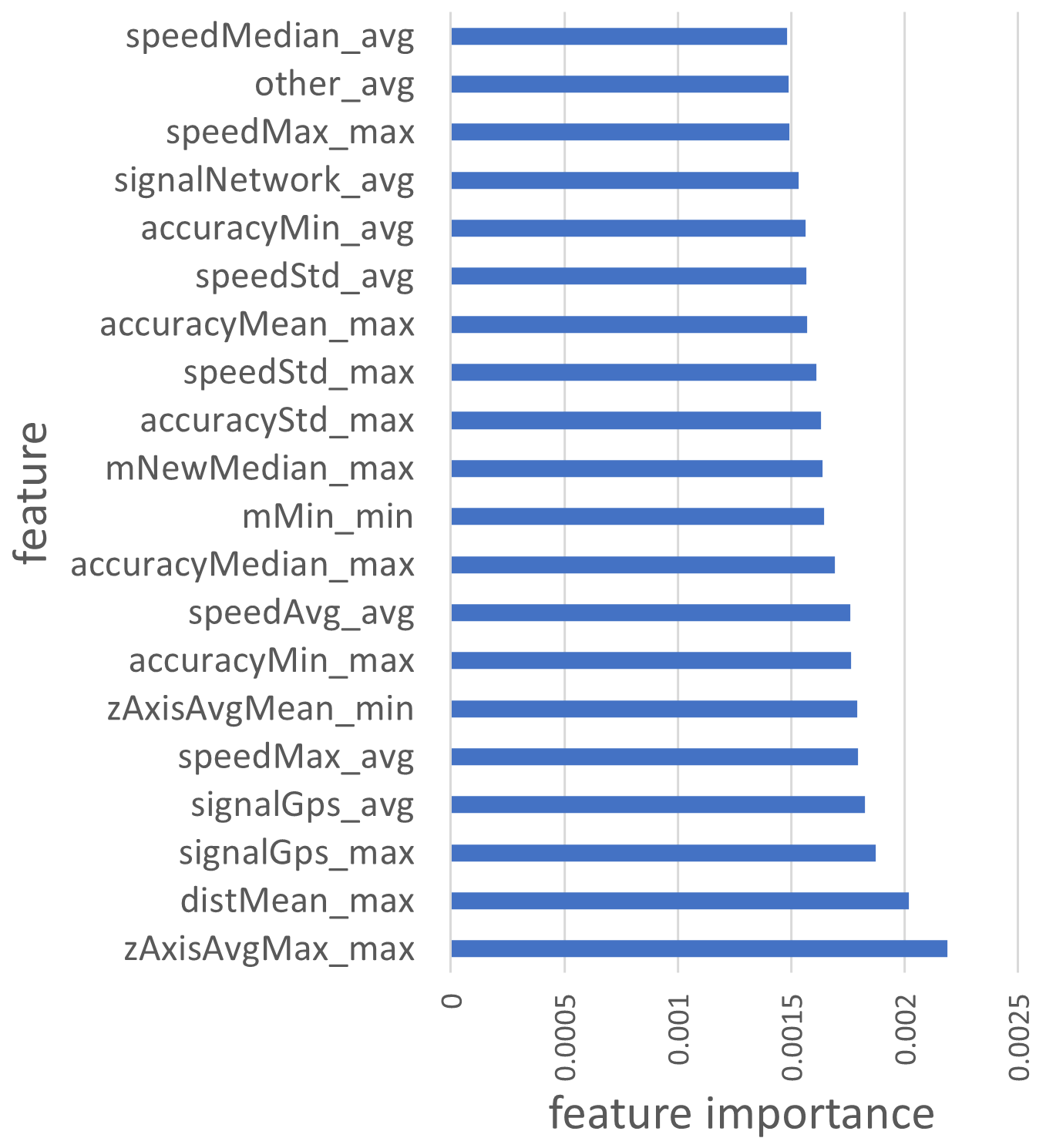}
        \caption{family$_{three}$}
        \label{fig:family_three}
    \end{subfigure}
    \hfill 
    \begin{subfigure}[t]{0.33\textwidth}
        \centering
        \includegraphics[width=0.78\textwidth]{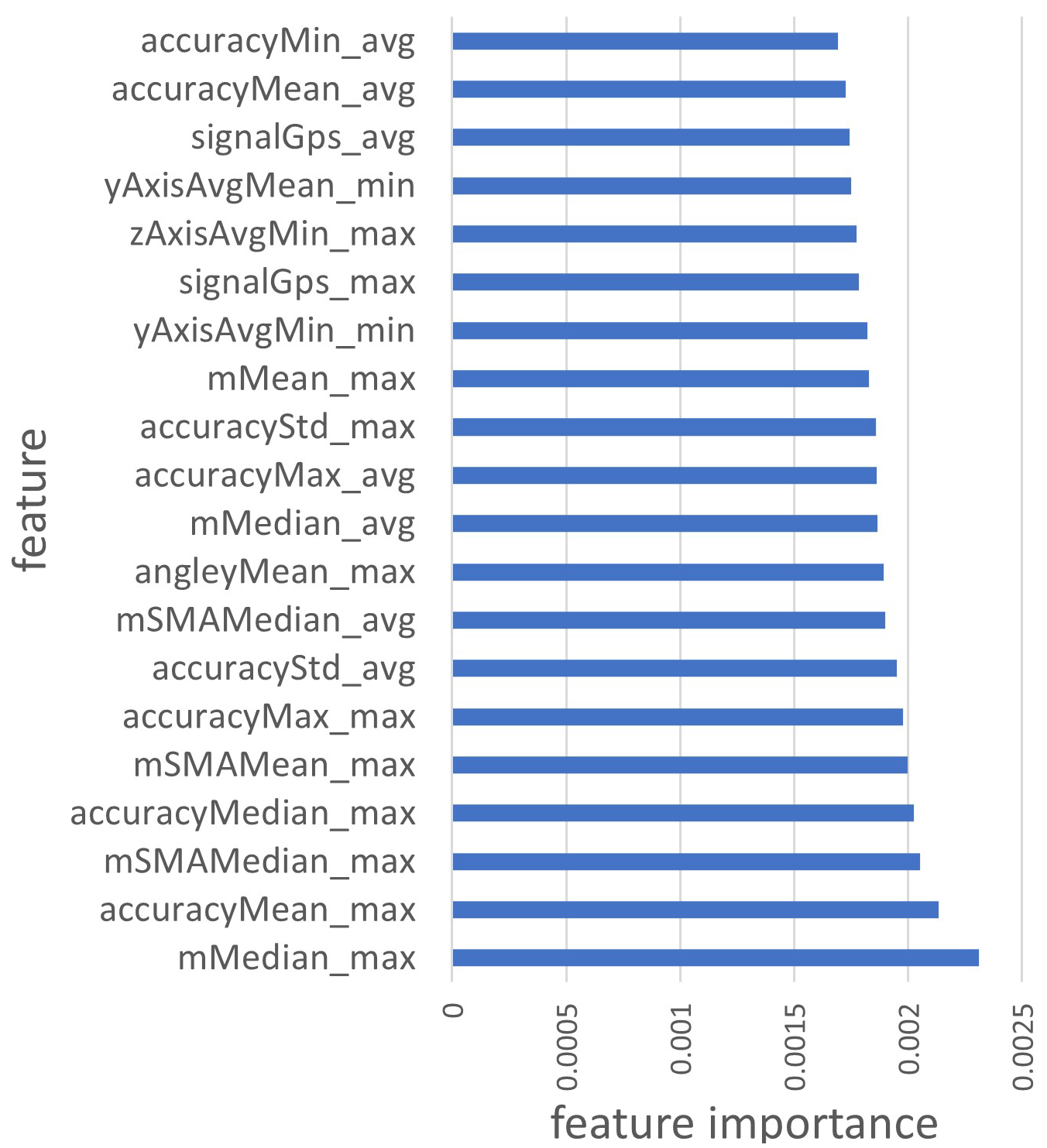}
        \caption{friends$_{three}$}
        \label{fig:friends_three}
    \end{subfigure}
    \hfill 
    \begin{subfigure}[t]{0.33\textwidth}
        \centering
        \includegraphics[width=0.78\textwidth]{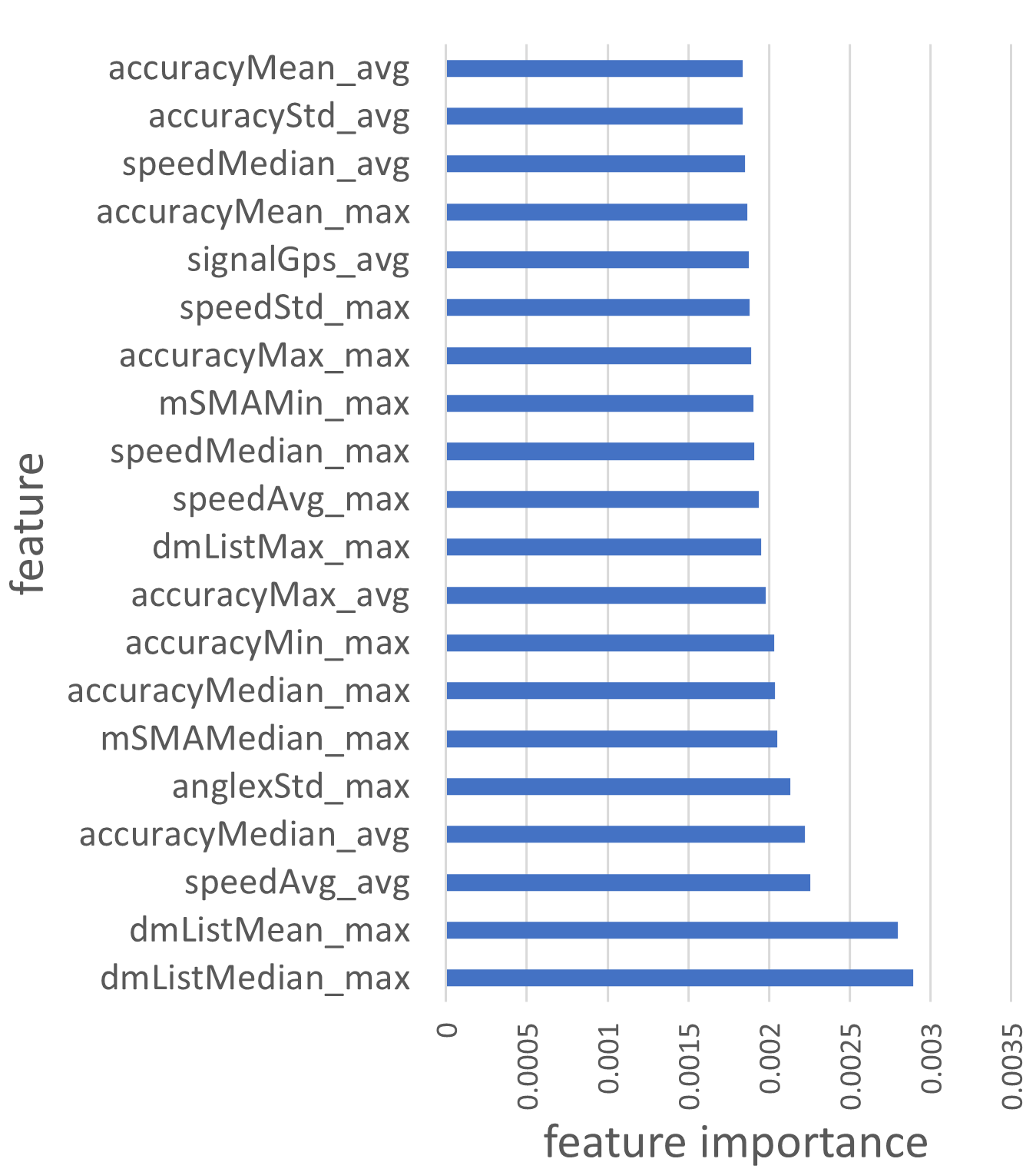}
        \caption{people$_{three}$}
        \label{fig:alone_three}
    \end{subfigure}

    \begin{subfigure}[t]{0.245\textwidth}
        \centering
        \includegraphics[width=1.08\textwidth]{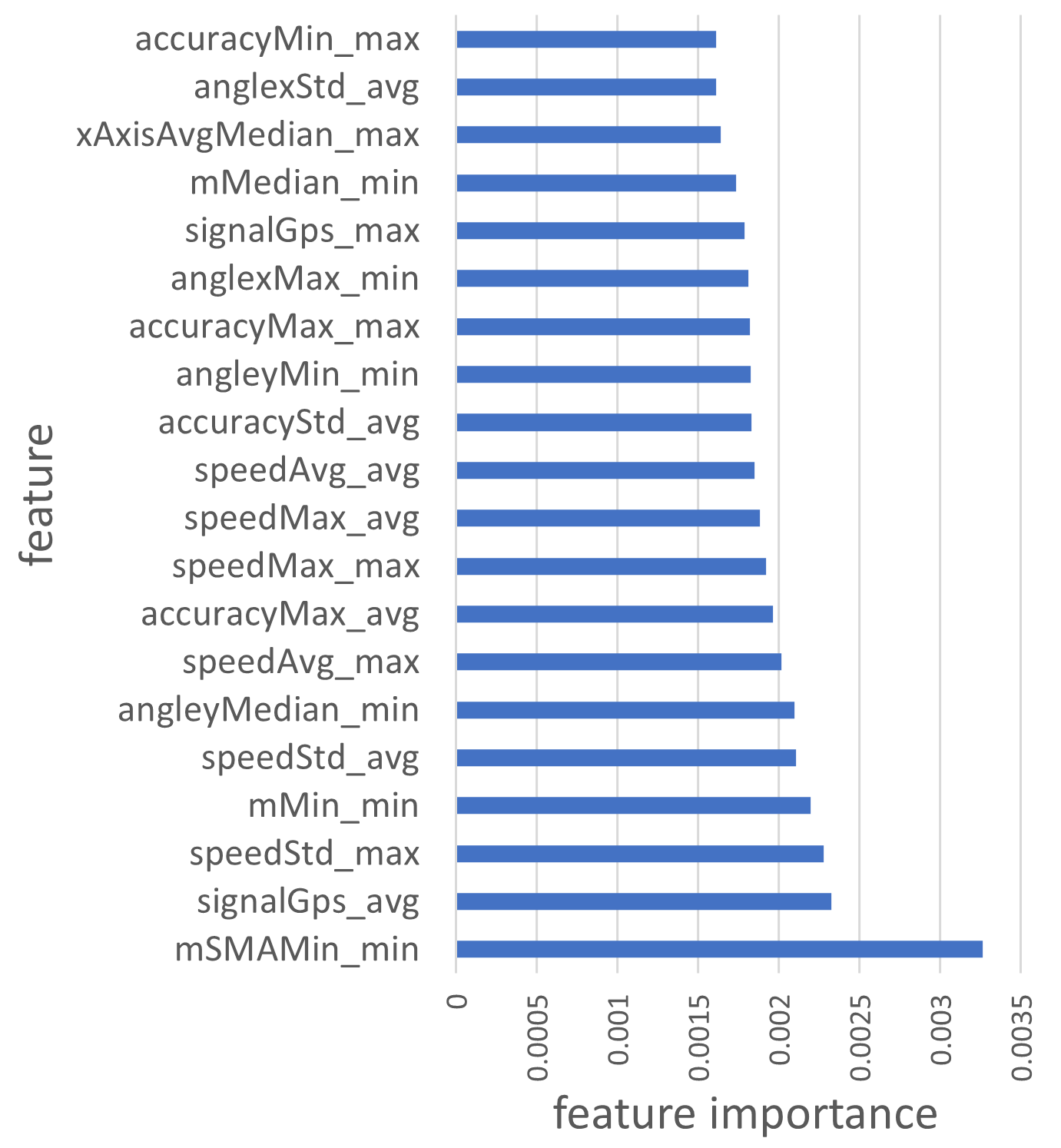}
        \caption{family$_{two}$}
        \label{fig:family_two}
    \end{subfigure}
    \hfill 
    \begin{subfigure}[t]{0.245\textwidth}
        \centering
        \includegraphics[width=1.08\textwidth]{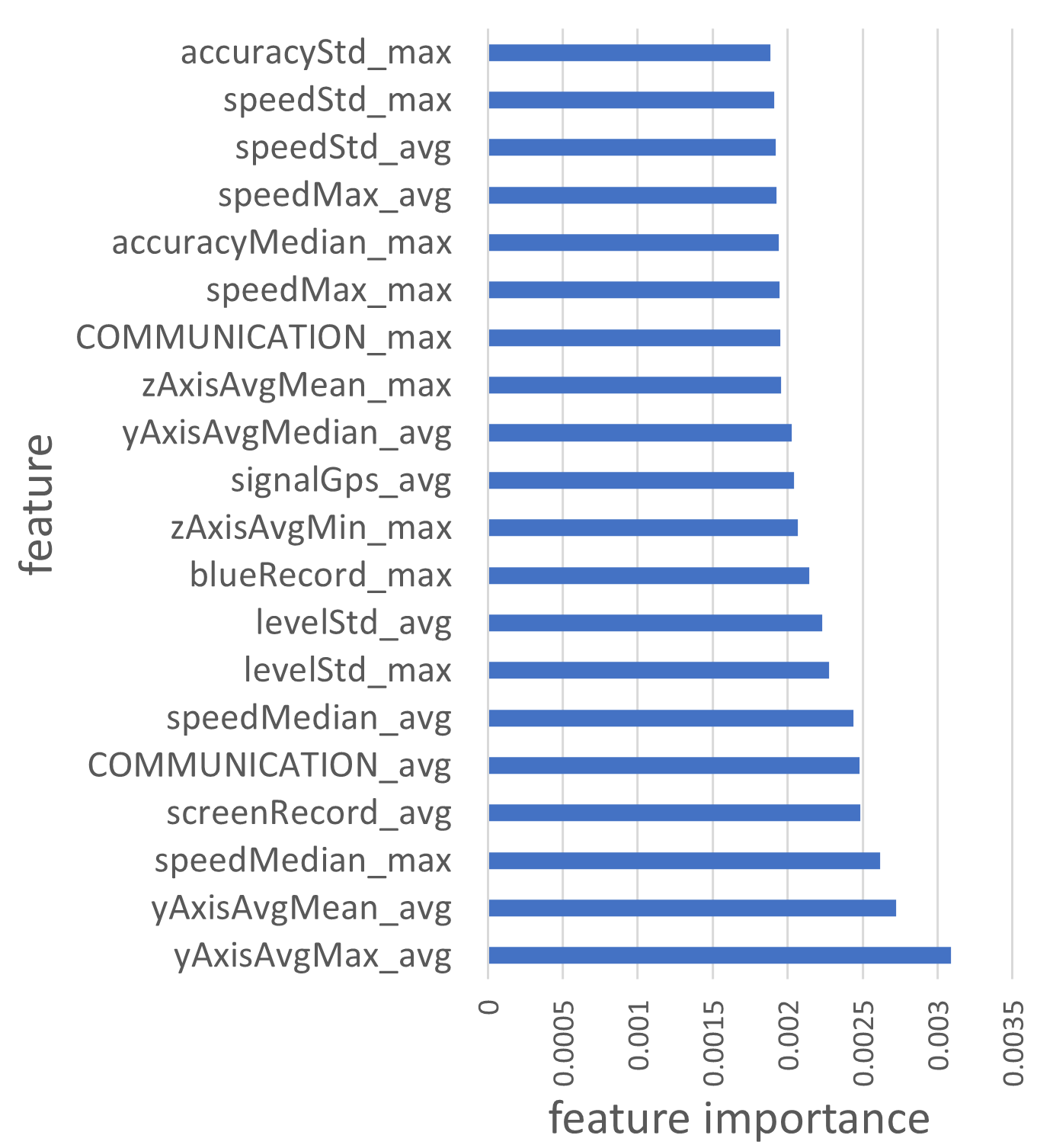}
        \caption{partner$_{two}$}
        \label{fig:partner_two}
    \end{subfigure}
    \hfill 
    \begin{subfigure}[t]{0.245\textwidth}
        \centering
        \includegraphics[width=1.08\textwidth]{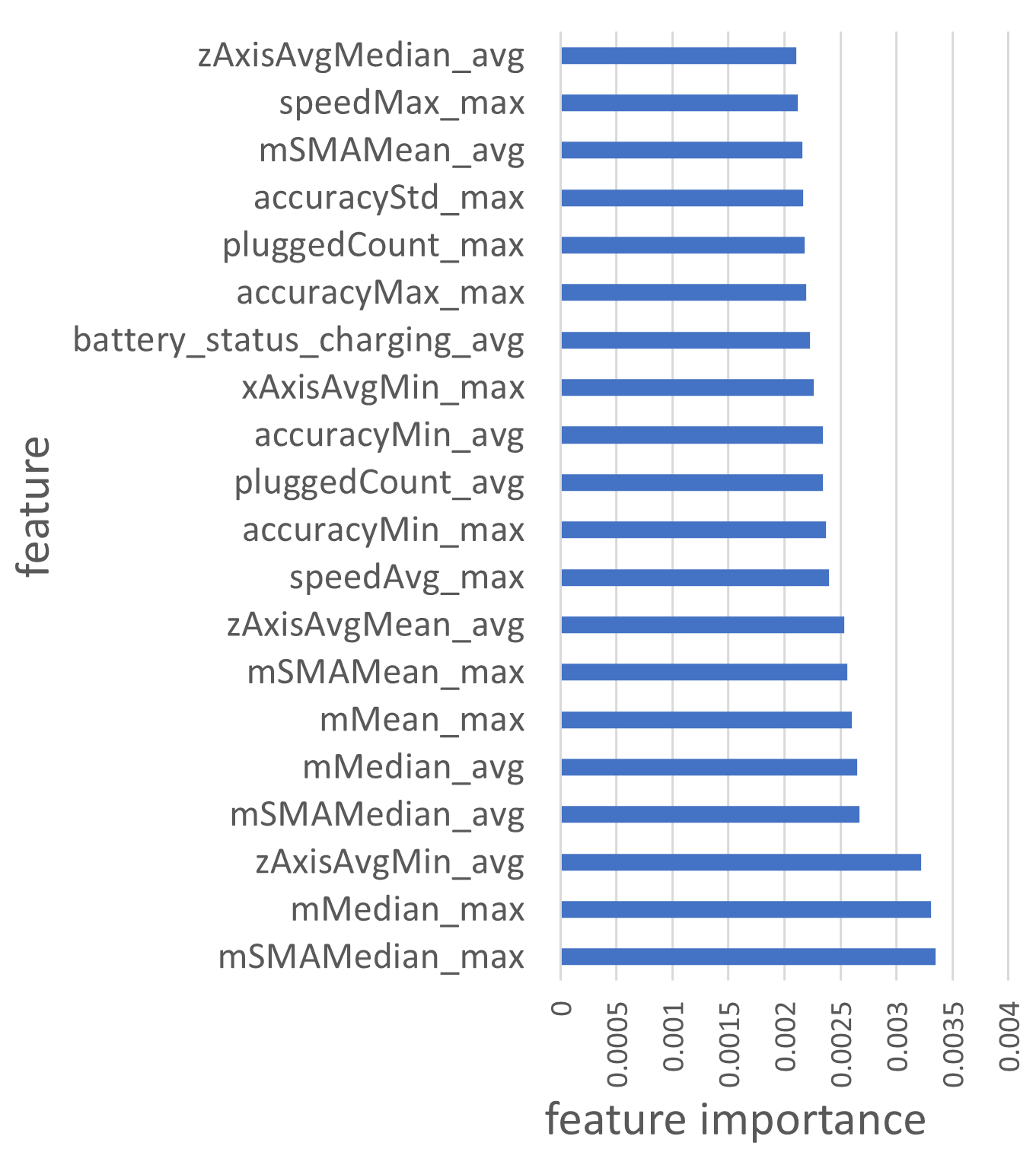}
        \caption{friends$_{two}$}
        \label{fig:friends_two}
    \end{subfigure}
    \hfill 
    \begin{subfigure}[t]{0.245\textwidth}
        \centering
        \includegraphics[width=1.08\textwidth]{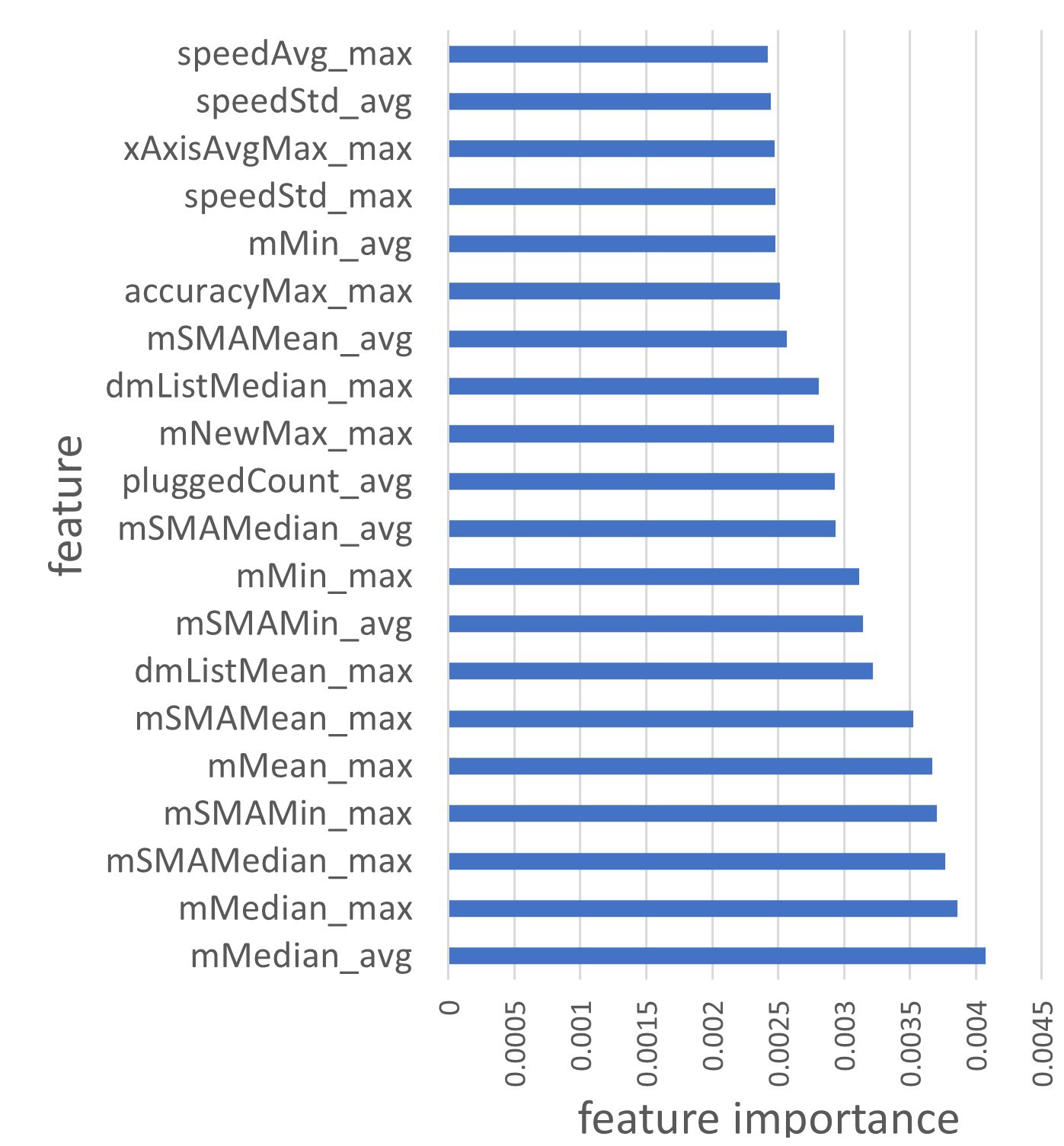}
        \caption{people$_{two}$}
        \label{fig:alone_two}
    \end{subfigure}
    
    \caption{Feature importance values from random forest classifiers with all features, for different social contexts.}
    \label{fig:feature_importance}
\end{center}
\vspace{-0.2 in}
\end{figure*}

\begin{figure*}[t]
\begin{center}
    \begin{minipage}[t]{\textwidth}
    \centering
        \includegraphics[width=\textwidth]{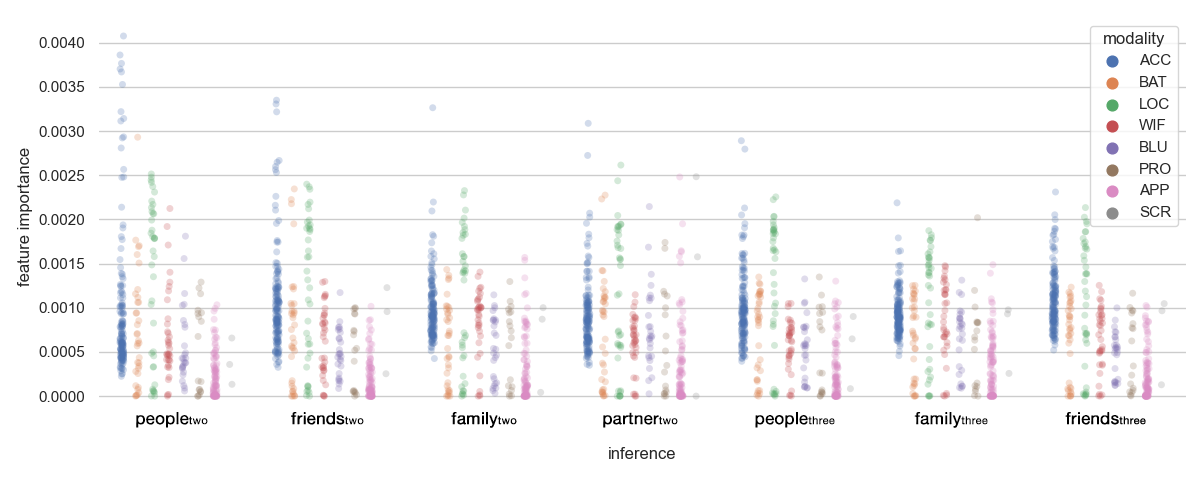}
        \caption{Feature importance value distributions for different sensing modalities in different inferences when using all features}
        \label{fig:feature_importance_distribution}
    \end{minipage}
\end{center}
\vspace{-0.2 in}
\end{figure*}

\subsection{Feature Importance for Social Context Inferences (RQ2)}\label{subsec:feature_importance}

In Figure~\ref{fig:feature_importance}, we show the top twenty feature importance values for each inference presented in Section 6.2. These values were captured from the output of trained random forest models, when using all features. We obtained feature importance values for all features in each iteration, and report the mean value for each feature. The sensor modality that was present throughout all seven inferences was the accelerometer (ACC). This is congruent with the results presented in the statistical analysis (Section~\ref{sec:statistical_analysis}). This suggests that physical activity levels and phone motion dynamics could help infer different types of social contexts of drinking occasions. This makes sense for certain situations because it is highly unlikely that a person would drink and dance alone on a weekend night, while this might happen when people are in larger groups (with both family and friends).

The second most common modality across all inferences is location (LOC). Features that capture the speed of the phone (speedMedian\_avg, speedMax\_avg, etc.), accuracy of the signal (accuracyMin\_max, accuracyMean\_avg, etc.), and signal type and strength (signalGps\_max, signalNetwork\_avg, etc.) are present across all inferences. Specially, for both two-class and three-class inferences regarding family, location features regarding GPS signal strength and speed filled up a majority of top five features. This suggests that location-related features have captured certain differences with regards to group dynamics in the social context family. Even though interaction sensing modalities (APP, SCR) were not present among all the social contexts, partner$_{two}$ had several features (COMMUNICATION\_avg, COMMUNICATION\_max, etc.) regarding communication app usage (e.g. viber, whatsapp, messenger, etc.) and also screen usage (screenRecord\_avg). Given interaction sensing modalities capture the phone usage behavior, this suggests that people use their phone differently when they are drinking alcohol with their partner as opposed to not being with him or her. 

In Figure~\ref{fig:feature_importance_distribution}, we plot a distribution of feature importance values for all social context inferences, grouped by different sensing modalities. This provides an overview of the informativeness of sensing modalities in making inferences. The most sparse distribution across all inferences came from the ACC, for the social context people$_{two}$. Overall, accelerometer produced the most informative feature, for all seven social contexts. Location features had comparatively high values for all seven social contexts. Even though location features were not among the highest for any inference, mean feature importance for location modalities was even higher than for accelerometer features (because the location feature distribution is negatively skewed). In addition, except for WIF, all other modalities had comparatively sparse and wider distributions for the context partner$_{two}$. To sum up, the takeaways from this analysis are: accelerometer features (ACC) were informative for all inferences, location features (LOC) were generally informative across all inferences too, application usage (APP) and screen usage (SCR) features (interaction sensing) were informative for partner$_{two}$ while not being comparatively informative for other inferences, and except for Wifi features (WIF), all other features had wider distributions for partner$_{two}$.

\subsection{Effect of Varying Group Sizes (RQ3)}\label{subsec:group_sizes}

In the previous analyses, we considered group dynamics as follows: (a) two-class social contexts - \textit{without} vs. \textit{with one/more people} and (b) three-class social contexts - \textit{without} vs. \textit{with one person} vs. \textit{with two/more people}. Hence, while the two-class inference mostly relate to the absence of a particular type of people, the three classes effectively tried to infer the presence of groups of varying sizes (with one person, with two/more people). If we consider the three-class inferences, \textit{with one person} means that it is a group of two people including the participant, and \textit{with two/more people} means that the group has a minimum size of three people including the participant, hence both classes capture  different group sizes, with the former being a small group, and the latter being a larger group in comparison. Given that there is no gold standard regarding the definition of the size of the drinking group (as highlighted in Section~\ref{sub:related_work}), in this section, we aim to change the size of these two groups by changing the threshold called \textit{grouping threshold}, which was always equal to \textit{one} in previous sections (e.g., without vs. with \textit{one}/less people vs. with two/more people), for three-class inferences. To this aim, we increase the value of the grouping threshold from one to ten, to investigate how it affects the inference accuracy. One and ten were chosen as highest and lowest thresholds because those were the highest and lowest values available in self-reports to define three-classes.

\begin{wrapfigure}{}{0.45\textwidth}
  \begin{center}
    \centering
        \includegraphics[width=0.45\textwidth]{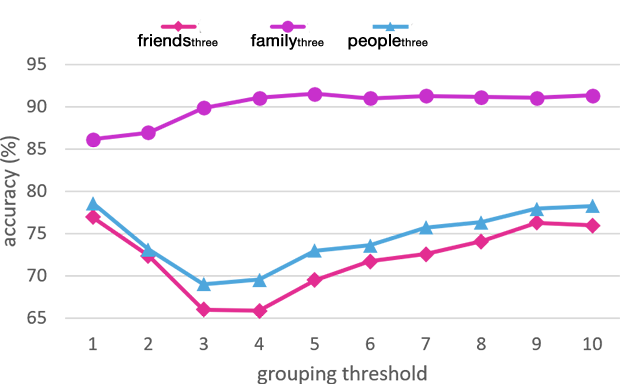}
        \caption{Three-Class Social Context Inference Accuracies for Different Grouping Thresholds}
        \label{fig:grouping}
  \end{center}
  \vspace{-0.2 in}
\end{wrapfigure}

We conducted the evaluation with the three three-class inferences using the same approach mentioned in Section~\ref{subsec:inference}, and the results are summarized in Figure~\ref{fig:grouping}. For friend$_{three}$ and people$_{three}$, inference accuracies decreased when increasing the grouping threshold, meaning that the model was not good at discerning the three classes when the threshold was around three (alone vs. with three/less people vs. with four/more people) and four (alone vs. with four/less people vs. with five/more people). However, when increasing the threshold further, the accuracies increased back to the same level as when the threshold was equal to one. What this means is that random forest classifier is not performing well when the small and large groups are defined by thresholds of the range three to five. This result is not surprising because any kind of nightlife-related activities available for a small group of people (they might find a table to fit altogether in a pub or a restaurant, they might easily travel with a cab, they might all gather in a living room) would result in a large heterogeneity of sensor data as compared to a larger group (e.g. ten or more people). This is because of the differences in behavior when people are in large groups, as opposed to small groups. Consequently, this would result in a lower inference accuracy when social contexts with three, four, or five people are in both the small group and the large group classes of the three-class inference. Consider an example where the grouping threshold is three, where samples with a group of three people would fall into the small group and samples with a group of four or more people would be included in the large group of the three-class inference. According to the distribution given in Figure~\ref{fig:dist_ori}, for the variable friends$_{three}$, when 114 samples (group of 3) and 105 samples (group of 4) fall into small and large groups in the inference, both classes have homogeneous sensor data, hence making it difficult for the model to discriminate between the classes. Conversely, the range of activities is smaller for larger groups due to its size, resulting in a lower heterogeneity of sensor data within groups, and consequently, higher inference accuracy for higher grouping thresholds. For example, consider the grouping threshold of ten where the small group would have ten or less people and the large group would have eleven or more people. According to Figure~\ref{fig:dist_ori}, for the variable people$_{three}$, there are 221 samples of eleven or more (clearly a large group), and over 300 samples of groups with three to ten people, with a majority of data coming from small group sizes such as three (135 samples), four (93 samples), and five (119 samples). This leads to heterogeneous data between small and large groups, because small group consists of data predominantly from groups of three, four, or five, and the large group is predominantly containing groups of 10+ people. This makes it easier for the model to discriminate between the three classes, hence leading to higher accuracies. On the other hand, for family$_{three}$, increasing threshold had an opposite effect, and increased the performance of the models. Again, this might be explained by the lower diversity of choices of activities and contexts to be sensed in family contexts, that tend to be highly routinized. Finally, results suggest that, regardless of the grouping threshold, models performed reasonably well for all family$_{three}$ inferences. In addition, except for grouping thresholds from three to five, for all other threshold, friends$_{three}$ and people$_{three}$ showed reasonable performance with accuracies over 70\%. Hence, according to this analysis, having different grouping thresholds seems a valid design choice depending on the application and the use-case.

\subsection{Sex Composition of Groups of Friends (RQ3)}\label{subsec:same_opposite_sex}

In the related work section, we described the importance of identifying the gender composition of people in groups when consuming alcohol. For example, we described how prior work discussed about men or women feeling more comfortable when drinking with groups of same sex friends \cite{Thrul2017,ander_2017}. In this section, we define and evaluate a three-class inference task for drinking episodes that are done with friends/colleagues (N = 799), with the classes: same-sex (389), opposite-sex (97), and mixed-sex (313). This feature was derived using the demographic sex attribute of the participant and the men and women friends/colleagues present in the drinking occasion, as reported by the participants. We followed the same approach as in Section~\ref{subsec:inference} to conduct the evaluation, and the results are presented in Table~\ref{tab:inference_sex}. According to results, the random forest classifier performed the best with an accuracy of 75.86\%, followed by gradient boosting that had an accuracy of 71.57\%. The ten highest feature importance values for the inference that were obtained using the random forest classifiers included six features from LOC (related to GPS signal strength and the accuracy of the signal, e.g. accuracyMedian\_max, accuracyMax\_max, etc.) and four from the ACC (reading from the z axis, e.g. zAxisAvgMedian\_avg, zAxisAvgMin\_avg; and aggregated m statistic - mMedian\_max). This result suggests that mobile sensing features can to some degree classify the sex composition of drinking groups, but more in-depth work would be needed to understand this phenomenon.

\begin{table}[t]
        \small
        \centering
        \caption{Sex-Composition Inference results for alcohol consumption episodes done with friends, in the order: \={A} (A$_{\sigma}$), AUC }
        \resizebox{0.85\textwidth}{!}{%
        \begin{tabular}{ l l l l l l}

        \rowcolor{red!5}
        
        \textbf{Target Variable}& 
        \textbf{Random Forest} &
        \textbf{XG Boost} &
        \textbf{Ada Boost} &
        \textbf{Gradient Boost}&
        \textbf{Naive Bayes} 
        \\ \hline 
        
        baseline &
        33.3 (0.0), 50.0 & 
        33.3 (0.0), 50.0 & 
        33.3 (0.0), 50.0 & 
        33.3 (0.0), 50.0 & 
        33.3 (0.0), 50.0 
        \\
        
        \rowcolor{gray!5}
        sex\_composition$_{three}$  &
        75.8 (2.5), 74.9 & 
        69.6 (4.1), 70.7 & 
        66.1 (6.1), 66.2 & 
        71.5 (3.7), 70.9 & 
        58.7 (7.3), 59.9 
        \\ \hline 
        
        \end{tabular}}
        \label{tab:inference_sex}

\end{table}

\section{Discussion}\label{sec:discussion}

\noindent \textbf{Features: }\label{subsec:modelpersonalization}It is worth noting that for modalities such as ACC and LOC, we generated simple statistical features that do not need extensive processing of the dataset. If we consider the ACC, while features proved to be informative in inferring different social contexts, the only set of features we used are statistical features from the three axes, angles between the gravity vector and axes, and aggregate features that combine the values of three-axes (Section~\ref{subsec:passive_sensing}). It is also worth noting that these feature are less interpretable in the context of alcohol consumption. For example, the feature mMedian\_max had the highest feature importance value for friends$_{three}$, as shown in Figure~\ref{fig:family_three}. While this feature represents the overall acceleration of the phone at a time period closer to the drinking event, it is difficult to interpret it compared to more interpretable features such as step count or activity type. If such features were derived using the accelerometer data, the interpretation could have been much simpler. However, we were not able to derive them due to limitations in the original dataset (sampling frequency, lack of gyroscope data, etc.). Future work could consider using low-power consuming libraries such as Google Activity Recognition API to obtain activity types and native step counters available in modern smartphones to obtain step counts, hence obtaining more interpretable features. In addition, researchers could also look into using other sensing modalities such as ambient light sensor, typing and touch events, and notification clicking behaviors. 

\noindent \textbf{Ethical Considerations: }\label{subsec:ethics} The goal of this study is to support public health research. Hence, it is essential to be aware of ethical implications. For public health, the inferences done in this work are anonymous in the sense that no identities of individuals are inferred when inferring social contexts. However, certain social contexts such as 'being with a partner' could be more sensitive, because identifying the presence of such people could potentially reveal sensitive information about them, even though they might not have agreed to have their location indirectly reported. Given that social context is relational, it is critical that during data collection, social companions (friends, family, etc.) agree that their presence is reported (even as an aggregate). Future studies should consider these aspects. Furthermore, for future interactive health systems that would be used by individuals and their health providers, it is fundamental to have clarity on who could access inferred data regarding social contexts, given their sensitive nature. Further, running social context inferences on-device, rather than on servers, would help preserve the privacy of users and others interacting with them. More generally, participants' respect of privacy and well-being should be the guiding lights of any future design of mobile health systems regarding alcohol consumption. 

\noindent \textbf{Importance of Diversity-Awareness:} The drinking behavior of people differ significantly depending on age, sex, drinking culture, beverage preferences, as well as how people perceive drinking alcohol \cite{Grittner2019, Bae2017, Santani2018}. For example, in some Asian countries, drinking alcohol might not be socially accepted while it is a societal norm in Europe and North America \cite{Sornpaisarn2020,Balhara2012}. Hence, it is worth pointing out that this study regarding the drinking behavior of young adults in Switzerland is exploratory, and the results cannot be directly assumed as being representative of the drinking behavior in other countries. Recent work has highlighted the importance of considering diversity-awareness when building social platforms using machine learning models and mobile sensing data \cite{Khwaja2019, Schelenz2021}.

\noindent \textbf{Limitations and Future Work:} We prepared the drinking event level dataset (in Section~\ref{sec:aggregation}) without assuming any relationship between two drinking events that occur consecutively, hence, we considered alcohol drinking events to be independent of each other. However, in reality, there could be a relationship between drinking events of the same person during the same night. Understanding such relationships is a complex problem, and it needs further examination. Another limitation of our work is that it does not capture complex relationships within family members. For example, young adults might prefer drinking with their brother or same-age members of the family, where as they might not feel comfortable drinking with their parents. In addition, the perception of parents and other family members could differ significantly, and it could affect the drinking behavior in the vicinity of family members. Furthermore, the partner's/spouse's perception towards alcohol consumption is another variable that was not captured during this study. These aspects need further investigation. In addition, it is worthwhile to note that, inferring the social context of drinking does not directly help overcome health problems. This is not the intention of this work, as it would oversimplify the problem. However, inferring the social context of drinking would assist or complement other inferences such as drinking occasions, drinking nights, drink vs. drunk in ubicomp and alcohol research \cite{Santani2018, Bae2017, Bae2018, Phan2019v2, Phan2020, Gustafson2014, Dulin2013}. In this respect, the inference of social context might help to provide meaningful and context-aware interventions that might decrease the amounts consumed, and, as a consequence, less adverse alcohol-related consequences. The design of such interventions is beyond the scope of the paper.

Another important aspect is the choice of time windows for aggregation and matching phases. Even though we presented results for the dataset obtained with ten-minute time window for aggregation and one-hour time window for matching, we conducted evaluations with different time windows. We obtained the best results using these time windows, and hence, considering space limitations, we only presented results for these windows. It is worth noting that the time window would affect the number of self-reports included in the study. For instance, if the matching time window is two hours, we need to discard all self-reports from 8.00pm to 9.00pm because we would not have enough sensor data for reports done between those time windows. The same applies to drinking events done between 3.00am and 4.00am. Further, it is worth noting that, regardless of the time window and the resulting dataset size, we obtained inference results comparable to the ones we presented in Table~\ref{tab:inference_results}, with differences of the range 0.4\% (best-case scenario) to 12\% (worst-case scenario). 

An important topic for future work is regarding the drinking motives of young adults. As we discussed in Section~\ref{sub:related_work}, drinking motives could be the primary driving factor why young adults choose specific social contexts to drink \cite{Kuntsche2009, Kuntsche2005}. Hence, examining the associations between such drinking motives and smartphone sensing data could further advocate the idea of building holistic mobile health systems that consider not only the alcohol consumption, but also other factors associated to the event. Furthermore, even though there were multiple comparisons in the statistical analysis, we did not use Bonferroni correction for p-values \cite{Vickerstaff2019}. Hence, the results with p-values should be interpreted with caution.

\section{Conclusion}\label{sec:conclusion}

In this study, we examined the weekend drinking behavior of 241 young adults in Switzerland using self-reports and passive smartphone sensing data. Our work emphasized the importance of understanding the social context of drinking, to obtain a holistic view regarding the alcohol consumption behavior. With multiple statistical analyses, we show that features from modalities such as accelerometer, location, bluetooth, and application usage could be informative about social contexts of drinking. In addition, we define and evaluate seven inference tasks obtaining accuracies of the range 75\%-86\% in two-class and three-class tasks, showing the feasibility of using smartphone sensing to detect social contexts of drinking occasions. We believe these findings could be useful for ubicomp and alcohol epidemiology researchers towards in implementing future mobile health systems with interventions and feedback mechanisms.

\begin{acks}

This work was supported by the Swiss National Science Foundation (SNSF) through the Dusk2Dawn project (Sinergia program) under grant number CRSII5\_173696.

\end{acks}

\bibliographystyle{ACM-Reference-Format}
\bibliography{citations}

\end{document}